\documentclass[twocolumn,trackchanges]{aastex7}
\usepackage{amsmath}

\usepackage{ulem}

\usepackage{float}
\usepackage{tabularx}
\usepackage{booktabs}
\usepackage{placeins}
\usepackage{verbatim}

\graphicspath{{./}{figures/}}
\usepackage{color} 
\makeatletter
\newcommand*{\chk}{\@ifnextchar\bgroup{\chk@}{\color{blue}}}
\newcommand*{\chk@}[1]{{\textcolor{blue}{#1}}}

\begin{document}

\title{The Salamander: A case study of the magnetic field and peculiar morphology of G309.8$-$2.6 through radio polarimetry}

\author[0009-0004-4090-9304]{Wenhui Jing}
\affiliation{School of Physics and Astronomy, Yunnan University, Kunming, 650091, P. R. China}
\email[show]{wh.jing@outlook.com}

\author[0000-0001-7722-8458]{Jennifer L. West}
\affiliation{Dominion Radio Astrophysical Observatory, Herzberg Astronomy \& Astrophysics, National Research Council Canada, P.O. Box 248, Penticton, BC V2A 6J9, Canada}
\email[show]{jennifer.west@nrc-cnrc.gc.ca}

\author[0000-0002-3464-5128]{Xiaohui Sun}
\affiliation{School of Physics and Astronomy, Yunnan University, Kunming, 650091, P. R. China}
\email[show]{xhsun@ynu.edu.cn}

\author[0000-0001-5953-0100]{Roland Kothes}
\affiliation{Dominion Radio Astrophysical Observatory, Herzberg Astronomy \& Astrophysics, National Research Council Canada, P.O. Box 248, Penticton, BC V2A 6J9, Canada}
\email{}

\author[0000-0002-4400-4703]{Isabel Sander}
\affiliation{Department of Physics and Astronomy, University of Manitoba, Winnipeg, MB R3T 2N2, Canada}
\email{}
\author[0000-0001-6189-7665]{Samar Safi-Harb}
\affiliation{Department of Physics and Astronomy, University of Manitoba, Winnipeg, MB R3T 2N2, Canada}
\email{}
\author[0000-0002-4814-958X]{Denis Leahy}
\affiliation{Department of Physics and Astronomy, University of Calgary, Calgary, AB, T2N 1N4, Canada}
\email{}
\author[0000-0002-3382-9558]{B. M. Gaensler}
\affiliation{Department of Astronomy and Astrophysics, University of California Santa Cruz, 1156 High Street, Santa Cruz, CA 95064, USA}
\affiliation{Dunlap Institute for Astronomy and Astrophysics, University of Toronto, 50 St. George Street, Toronto, ON M5S 3H4, Canada}
\affiliation{David A. Dunlap Department of Astronomy and Astrophysics, University of Toronto, 50 St. George Street, Toronto, ON M5S 3H4, Canada}
\email{}
\author[0000-0022-9399-8433]{Xianghua Li}
\affiliation{School of Physics and Astronomy, Yunnan University, Kunming, 650091, P. R. China}
\email[show]{xhli@ynu.edu.cn}
\author[0009-0003-2088-9433]{Brianna Ball}
\affiliation{Department of Physics, University of Alberta, Edmonton, Alberta, T6G 2E1, Canada}
\email{}
\author[0000-0002-6243-7879]{Craig Anderson}
\affiliation{Research School of Astronomy and Astrophysics, Australian National University, ACT 2611, Australia}
\email{}
\author[0000-0003-1173-6964]{W. Becker}
\affiliation{Max-Planck-Institut f\"ur extraterrestrische Physik, Giessenbachstra{\ss}e, 85741 Garching, Germany\\}
\affiliation{Max-Planck-Institut f\"ur Radioastronomie, Auf dem H\"ugel 69, 53121 Bonn, Germany}
\email{}
\author[0000-0002-4990-9288]{Miroslav D. Filipovi\'c}
\affiliation{Western Sydney University, Locked Bag 1797, Penrith South DC, NSW 2751, Australia}
\email{}
\author[0000-0002-6097-2747]{Andrew M. Hopkins}
\affiliation{School of Mathematical and Physical Sciences, 12 Wally’s Walk, Macquarie University, NSW 2109, Australia}
\email{}
\author[0000-0003-0742-2006]{Yik Ki Ma}
\affiliation{Max-Planck-Institut f\"ur Radioastronomie, Auf dem H\"ugel 69, 53121 Bonn, Germany}
\email{}
\author[0000-0003-2730-957X]{Naomi McClure-Griffiths}
\affiliation{Research School of Astronomy and Astrophysics, Australian National University, ACT 2611, Australia}
\email{}
\author[0000-0001-9414-175X]{Syed Faisal ur Rahman}
\affiliation{SBASSE at Lahore University of Management Sciences, LUMS, Lahore Pakistan}
\affiliation{NCBC at NED University of Engineering and Technology, Karachi Pakistan}
\email{}
\author[0000-0002-7641-9946]{Cameron L. Van Eck}
\affiliation{Research School of Astronomy \& Astrophysics, The Australian National University, Canberra, ACT 2611, Australia}
\email{}
\author[0000-0002-1272-3017]{Jacco Th. van Loon}
\affiliation{Lennard-Jones Laboratories, Keele University, ST5 5BG, UK}
\email{}
\author[0009-0001-5653-9481]{Jayde Willingham}
\affiliation{School of Mathematical and Physical Sciences, 12 Wally’s Walk, Macquarie University, NSW 2109, Australia}
\affiliation{Macquarie University Astrophysics and Space Technologies Research Centre, Sydney, NSW 2109, Australia}
\email{}

\begin{abstract}

Studying the interaction between core-collapse supernova remnants (SNRs) and their surrounding environments is essential to understanding the mechanism for energy transfer to the interstellar medium (ISM) and the intrinsic physical properties of these remnants. In this paper, we focus on G309.8$-$2.6. Our new observations reveal that this object includes an SNR shell with a relic pulsar wind nebula (PWN) that extends well beyond the emission that has been previously observed in X-rays. We present new radio continuum and polarization images of G309.8$-$2.6 from the Evolutionary Map of the Universe (EMU) and Polarization Sky Survey of the Universe’s Magnetism (POSSUM) surveys with the Australian Square Kilometre Array Pathfinder (ASKAP). The images reveal the complex and peculiar morphology of G309.8$-$2.6. The linear polarization displays an atypical S-shaped morphology and a highly ordered magnetic field. The rotation measure (RM) map shows a large-scale gradient or possible sign reversal, depending on the foreground RM. We reprocessed archival X-ray observations from \textit{Chandra} and \textit{eROSITA}, and retrieved archival $\mathrm{H}\alpha$ and infrared observations. We performed a joint analysis of the multiwavelength data and proposed scenarios to explain the unusual shape. Our results place new constraints on the magnetic field of G309.8$-$2.6, including its environment, and demonstrate the power of polarization observations in probing the properties of SNRs.
\end{abstract}


\section{Introduction}\label{sec:introduction}

As a supernova remnant (SNR) evolves, any embedded pulsar wind nebula (PWN) will interact with the reverse shock from the remnant, leading to distorted PWN morphologies. In some cases, the pulsar escapes from the explosion site, leaving behind a ``relic'' PWN and forming a new compact nebula near its current position~\citep[see][for reviews]{Gaensler2006,Kothes2017,Olmi2023}. The relic PWN G309.8$-$2.6 provides a unique opportunity to understand how PWNe evolve and interact with their parent SNRs. 

\begin{figure}
    \centering

    \includegraphics[width=0.9\linewidth]{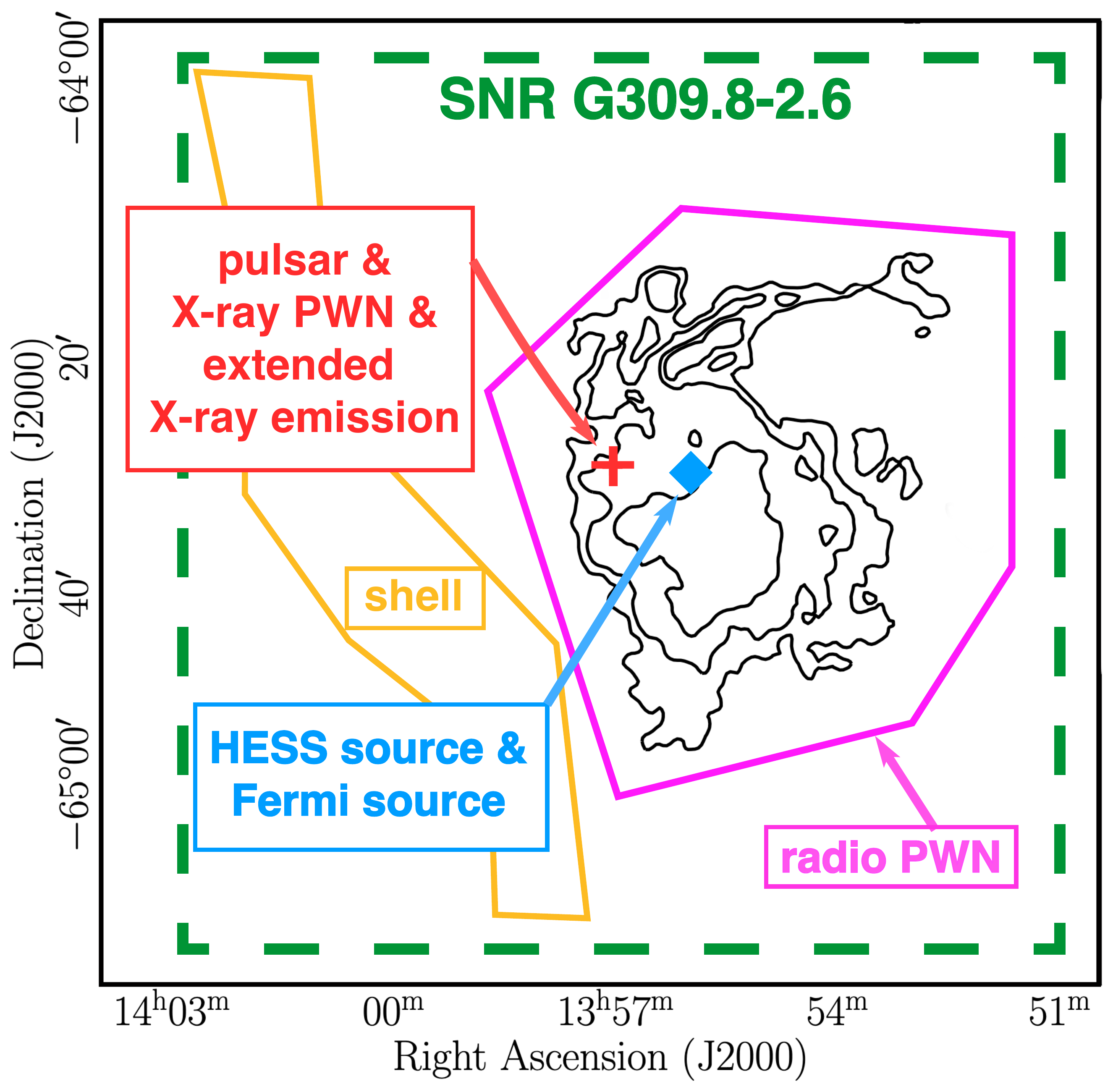}
    \caption{ Illustration of the main features of G309.8$-$2.6 mentioned in Section \ref{sec:introduction}. The contours are from the ASKAP total intensity. Note that the SNR here includes the relic PWN and the eastern shell outlined by the orange polygon. }
    \label{fig:features}
\end{figure}

G309.8$-$2.6 was initially identified as an SNR candidate by \cite{Duncan1997} in the Parkes 2.4 GHz survey~\citep{Duncan1995}, on the basis of its extended features and polarization signal. However, the object has many unusual characteristics, such as its lack of shell-type morphology and flat spectrum, indicating a potential classification as a PWN rather than a typical SNR~\citep{H.E.S.S.Collaboration2011}. 

The main features discussed above and below are displayed in Figure \ref{fig:features}, including the pulsar associated with an X-ray PWN and extended X-ray emission, the HESS and Fermi sources, and the extended PWN emission that is observed in radio and is offset from the X-ray PWN. 

On the eastern edge of the radio PWN, PSR J1357$-$6429 was discovered during the Parkes Multibeam Survey~\citep{Camilo2004}. 
Its pulsations have been detected in X-rays with \textit{XMM–Newton} and in GeV $\gamma$-rays with \textit{Fermi}~\citep{Esposito2007, Lemoine-Goumard2011, Chang2012}. The characteristic age of $\sim$7.3 kyr, combined with its spin-down luminosity of $\dot{E}$ = 3.1$\times$ 10$^{36}$ ergs s$^{-1}$, characterizes it as a young and energetic Vela-like pulsar. PSR J1357$-$6429 has a dispersion measure (DM) of $\sim$127 pc cm$^{-3}$, which places it at a distance of $\sim$3.1 kpc using the Galactic electron density model of YWM16 \citep{Yao2017}. While the previous distance estimate by \citet{Camilo2004}, based on the Galactic electron density model of \citet{Cordes2002}, places the source at a closer distance of $\sim$2.5\,kpc. This value is further supported by the optical extinction based distance of $2.3 \pm 0.2$\,kpc toward G309.8$-$2.6 reported by \citet{Shan2019a}. There are two previous measurements of the pulsar RM: $-$47$\pm$2\,rad\,m$^{-2}$~\citep{Johnston2006} and $-$43$\pm$1\,rad\,m$^{-2}$ \citep{Kirichenko2015}.

Observations with \textit{Chandra} and \textit{XMM-Newton} revealed a 2$\arcmin$ tail-like extended emission around PSR J1357$-$6429~\citep{Zavlin2007, Chang2012}. Spectral analysis suggests that this tail-like structure is an associated PWN, designated as PWN G309.92$-$2.51. This analysis also shows that the spectrum softens with increasing distance from the pulsar, although it is still unclear whether the PWN is part of a pulsar jet or of a pulsar tail~\citep{Chang2012}. No counterpart or tail-like structure at radio wavelengths has been reported. A broader field of view (FOV) X-ray study with \textit{ROSAT}, \textit{XMM-Newton}, and \textit{Suzaku} ~\citep{H.E.S.S.Collaboration2011,Izawa2015} revealed faint extended emission with a size of $\sim$$10\arcmin$. The peak of the extended emission is slightly offset from the pulsar and partially overlaps with the radio emission from G309.8$-$2.6. No significant spectral variation as a function of distance from the pulsar was found~\citep{Izawa2015}. \textit{NuSTAR} has observed this region~\citep{Mori2022}, with a detailed X-ray and multi-wavelength SED study forthcoming~\citep{Sander2026}.

One outstanding feature of the G309.8$-$2.6 field is the spatial connection with the TeV source HESS J1356$-$645~\citep{H.E.S.S.Collaboration2011}. \citet{Liu2023a} analyzed over 13 years of \textit{Fermi-LAT} data and detected significant GeV $\gamma$-ray emission. The GeV to TeV spectrum is well described by a leptonic model with a broken power law. The $\gamma$-ray properties of HESS J1356$-$645 resemble those of several known PWNe. However, because of limited spatial resolution, it cannot be determined whether the high-energy emission arises from a single source or from multiple components. Notably, its $\gamma$-ray emission varies with energy, which is similar to the Vela-X PWN, supporting a possible connection between these different emission regions~\citep{Liu2023a}.

The system comprising G309.8$-$2.6 and X-ray PWN G309.92$-$2.51 exemplifies a potential relic PWN scenario~\citep{Chang2012,Liu2023a,Sander2026}. A key parameter for understanding the evolution of extended emission regions is the energy loss rate of relativistic electrons, which is strongly influenced by the magnetic field. In radio observations, polarization measurements serve as a valuable tool for probing magnetic structures in complex systems. Because G309.8$-$2.6 is located relatively far from the Galactic plane, its polarization measurements are less affected by foreground contamination. The Parkes 2.4 GHz Survey~\citep{Duncan1997} shows polarized emission aligning with the northeastern region of the Stokes I candidate. This is consistent with the large-scale polarization, although the survey was constrained by limited angular resolution.

Recent observations from the Evolutionary Map of the Universe~\citep[EMU,][]{Hopkins2025} and Polarization Sky Survey of the Universe’s Magnetism~\citep[POSSUM,][]{Gaensler2025} surveys conducted by ASKAP show a highly polarized S-shaped source \citep{Ball2025}. The observation provides insights into the local magnetic field and raises questions about the origin of this system. 
Here we present a radio polarimetric and multiwavelength study of this source. The source has been named ``Salamander'' based on its X-ray emission near the pulsar PSR J1357$-$6325, which resembles a salamander (discussed in a forthcoming paper by \citealt[][]{Sander2026}). Its polarization morphology displays an S-shaped feature that also strongly resembles a salamander on a larger scale, and thus we use the name here to refer to the entire object observed in radio. In Section \ref{sec:data}, we describe the observations and data reduction procedures, while Section \ref{sec:results} provides the results. In Section \ref{sec:analysis}, we present a distance estimate and polarimetry analysis. Section \ref{sec:discussion} discusses the results, and Section \ref{sec:conclusion} provides the conclusions.

\section{Data acquisition and reprocessing}\label{sec:data}

We present new radio observations of G309.8$-$2.6 with ASKAP. We also reprocess or retrieve archival infrared, optical and X-ray observations for a multiwavelength analysis. Information of all the observations is listed in Table \ref{tab:multiwavelength}.

\begin{deluxetable*}{llllllll}
\tablecaption{ Multiwavelength observations used in the analysis\label{tab:multiwavelength}}
\tablehead{
\colhead{Wavelength} &
\colhead{Instrument} &
\colhead{Survey} &
\colhead{Band} &
\colhead{Resolution} &
\colhead{Figure} &
\colhead{Reference}
}
\startdata
Radio    & ASKAP  & EMU/POSSUM & 799--1087 MHz & $15\arcsec$ or $18\arcsec$ &   
Fig.~\ref{fig:SD-Int}, \ref{fig:radio-overall}, \ref{fig:relic-pi}, \ref{fig:pwn}, \ref{fig:shell}, \ref{fig:fdf}, \ref{fig:I-PI} & 1 \\
Radio    & Murriyang & STAPS & 1.3--1.8 GHz  & $20\arcmin$     & 
Fig.~\ref{fig:SD-Int}, \ref{fig:radio-overall} & 2 \\
Infrared & WISE   & --- & 12 $\mu$m & $6.5\arcsec$ & 
Fig.~\ref{fig:infrared-ha}, \ref{fig:shell} & 3 \\
Optical  & AAO/UKST & SuperCOSMOS & H$\alpha$ & $\sim0.5\arcsec$ & 
Fig.~\ref{fig:shell} & 4 \\
Optical  & --- & SHASSA & H$\alpha$ & $1\arcmin$ & 
Fig.~\ref{fig:infrared-ha} & 5 \\
X-ray    & Chandra & --- & 0.5--7.0 keV & $\sim0.5\arcsec$ & 
Fig.~\ref{fig:pwn} & 6 \\
X-ray    & eROSITA & eRASS & 0.5--7.0 keV & $\sim26\arcsec$ & 
Fig.~\ref{fig:erosita} & 7 \\
\enddata

\tablecomments{$^{1}$This paper; $^{2}$\citet{Sun2025}; $^{3}$\citet{Wright2010}; $^{4}$\citet{Parker2005}; $^{5}$\citet{Gaustad2001}; $^{6}$\citet{Chang2012}; $^{7}$\citet{Predehl2021, Merloni2024}.}

\end{deluxetable*}
\subsection{Radio}
\subsubsection{ASKAP}
\begin{deluxetable}{lc}

\tablecaption{Summary of ASKAP observations and image properties}
\label{tab:askap_obs}
\tablehead{
\colhead{Parameter} & \colhead{Value}
}
\startdata
Observation Date                  & 2023 September 30 \\
Center Frequency                  & 943 MHz \\
Bandwidth                         & 288 MHz \\
Resolution         & $15\arcsec$ for $I$ $18\arcsec$ for $P$ \\
RMSF FWHM                         & 59 rad m$^{-2}$ \\
$\sigma_I$ 
                                  & $25-30~\mu$Jy beam$^{-1}$ \\
$\sigma_P$  
                                  & $\sim13.5~\mu$Jy beam$^{-1}$ RMSF$^{-1}$ \\
\enddata

\end{deluxetable}
The field containing G309.8$-$2.6 is located near the center of the ASKAP scheduling block ID 53310~\citep{hopkins_askap_2022}. Observations were carried out for $\sim$10 hours on 2023 September 30 in the \textit{continuum averaged} mode. In this mode, the total 288~MHz bandwidth centered at 943~MHz is divided into 288 frequency channels, each 1~MHz wide. ASKAP simultaneously formed 36 independent beams in the standard ``closepack36'' configuration, covering $\rm \sim$$\rm 30~deg^2$~\citep{Hotan2021}.

We obtain the calibrated visibilities and image products from the CSIRO ASKAP Science Data Archive (CASDA)\footnote{\url{https://research.csiro.au/casda/}}. The available products include multi-frequency synthesis (MFS) mosaicked images of Stokes~$I$ and~$V$, as well as 1\,MHz resolution image cubes for all four Stokes parameters ($I$, $Q$, $U$, and $V$). The mosaicked images are convolved to a uniform resolution of 15$\arcsec$ for all beams, while the resolution of the spectral cubes varies with frequency, reaching up to 18$\arcsec$ at the lowest frequencies. For spectral and polarization analysis, we convolve all cubes to a common resolution of 18$\arcsec$. 

To improve sensitivity to extended structures, the full 288\,MHz bandwidth is re-binned into coarser sub-bands of both 144\,MHz and 72\,MHz for spectral analysis and 18\,MHz for polarization studies. For each case, we image and mosaic the nine beams covering the G309.8$-$2.6 region and convolve all resulting sub-band images to a common resolution of 18$\arcsec$.

\subsubsection{Combining Parkes and ASKAP maps}
The largest angular size of the ASKAP observation is 25$\arcmin$$-$50$\arcmin$ across the frequency range~\citep{McConnell2020a}. This implies that any smooth structure in our image with an angular size larger than this will be only partially detected. This limitation may be mitigated by combining the interferometer image with single-dish observations at the same frequency, ensuring an accurate representation of all structures from the largest scales down to the resolution limit~\citep{Plunkett2023}. Since no single-dish observations are available at the same frequency, short spacing data in Stokes $I$ are derived using a power-law extrapolation to 943\,MHz with the closest match from the Southern Twenty-centimetre All-sky Polarization Survey (STAPS) in the frequency range 1.3$-$1.8\,GHz~\citep{Sun2025}.  For the extrapolation, the power law spectral index $\alpha$, defined as $S_\nu\propto\nu^\alpha$, with $S_\nu$ being the flux density at frequency $\nu$, was obtained from fitting the STAPS data. The uncertainty of the extrapolated flux density at 943 MHz is $\sim2$\%.

\begin{figure*}
  \centering
  \includegraphics[width=1\linewidth]{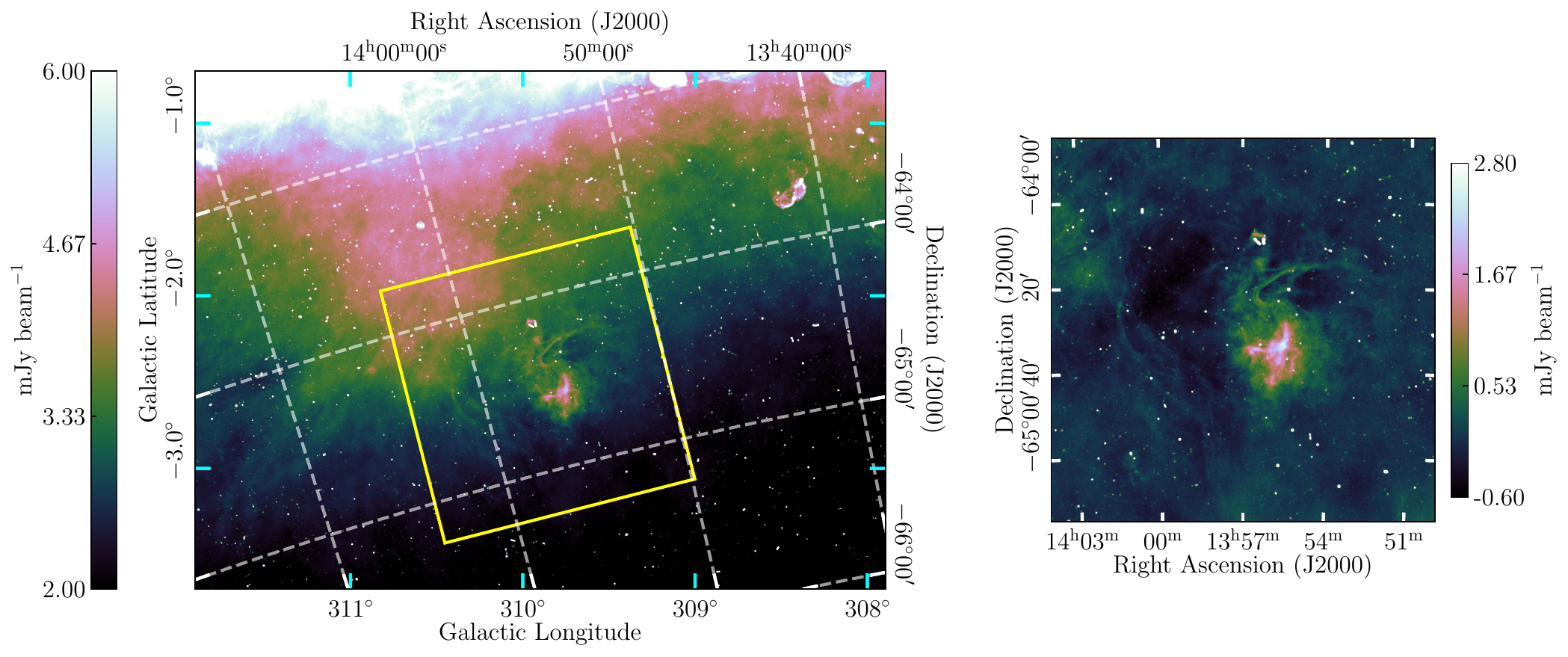}
  \caption{Left panel: The combined total intensity image from EMU and STAPS over a $4\degr \times 3\degr$ region. Right panel: The total intensity image from EMU for the $1.5\degr \times 1.5\degr$ region outlined with the yellow square in the left panel. Both images are at 943\,MHz with an angular resolution of $18\arcsec$, and displayed in cubehelix colour scheme \citep{Green2011}.}
  \label{fig:SD-Int}
\end{figure*}

\begin{figure}
  \centering
  \includegraphics[width=0.95\linewidth]{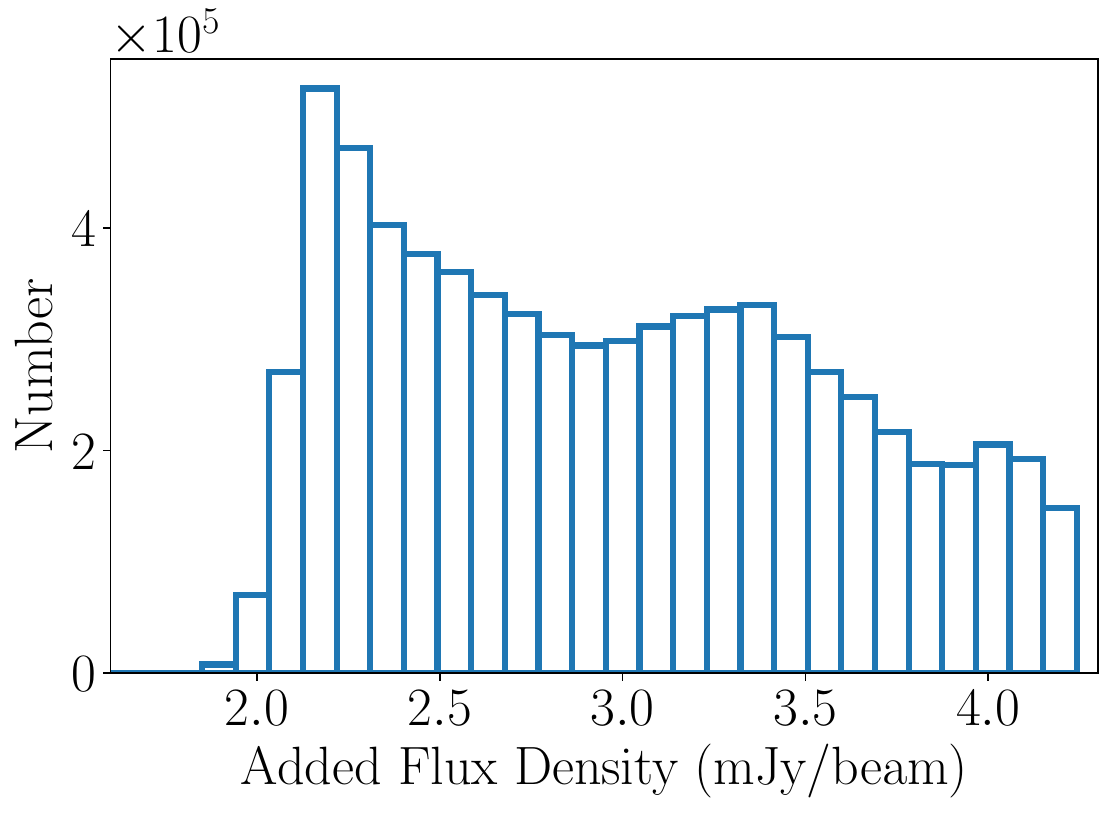}
  \caption{ Histogram of added flux densities after combining EMU and STAPS images.}
  \label{fig:hist}
\end{figure}

We apply the \texttt{feather} function in CASA \citep{CASA2022} to combine the Parkes and ASKAP data, ensuring a consistent flux scale across overlapping spatial frequencies~\citep[see][for a more detailed explanation]{Plunkett2023}. Additionally, we test the short-spacing correction method of \cite{Faridani2018} in the image domain. These two approaches yield similar results, with intensity levels for G309.8$-$2.6 being consistent to within 1\%, and we adopt the \texttt{feather} combination for the following analysis. The comparison of the original EMU observation image and the combined image is shown in Figure \ref{fig:SD-Int}. After combination, the added flux densities range from about 2 mJy beam$^{-1}$ to 4 mJy beam$^{-1}$ (Figure \ref{fig:hist}).

The combined image contains both the Galactic background extended emission ($I_{bkg}$) and the emission from G309.8$-$2.6 itself ($I_{\rm snr}$). We use \texttt{Aegean}~\citep{Hancock2012,Hancock2018} for fitting and subtracting the background and point-like sources. The final image in total intensity at 943 MHz is shown in Figure \ref{fig:radio-overall}. Note that the point source subtraction procedure was not able to remove all the sources, due to the extensive diffuse emission in this region.

The same procedure cannot be applied to Stokes $Q$ and $U$. Due to Faraday rotation and depolarization, the large-scale polarized emission from STAPS at a higher frequency and from ASKAP at a lower frequency are different. Therefore, the extrapolation from STAPS to ASKAP, as has been done for $I$, is not reliable for $Q$ and $U$. However, since $Q$ and $U$ generally show more small-scale variations and complex structures than Stokes $I$, this limitation may not be a significant issue. 
\subsubsection{RM synthesis}

The total linear polarized intensity \( P = \sqrt{Q^2+U^2}\) and the polarization angle \( \psi = \frac{1}{2}\text{arctan(}U/Q\text{)}\) are fundamental quantities in the study of magnetized astrophysical plasmas. The polarization angle is rotated due to the Faraday effect, which depends on the Faraday rotation measure (RM) as $\psi(\lambda^2) = \psi_0 + \text{RM} \lambda^2$, 
where \( \psi_0 \) is the intrinsic polarization angle, and \( \lambda \) is the wavelength. 
A more general term for RM is the Faraday depth, \(\phi(l)\), which can be written as $\phi(l) = 0.81 \int_{0}^{l} n_e(l^\prime) B_{\parallel}(l^\prime) {\rm d}l^{\prime}~\text{rad m\(^{-2}\)}$, where the integration is performed along the line of sight from the source to the observer, with $ l $ measured in pc. Here,  $n_e$ (in cm$^{-3}$) is the thermal electron density, and $B_{\parallel}$ (in $\rm \mu G$) is the line-of-sight component of the magnetic field. When the electron-density-weighted line-of-sight averaged magnetic field points toward the observer, $\phi$ is positive; otherwise, it is negative.

As shown by \citet{Burn1966} and \citet{Brentjens2005}, the observed complex polarized intensity $\mathcal{P}(\lambda^2)=Q(\lambda^2)+iU(\lambda^2)$ as a function of $\lambda^2$ and the complex polarized intensity $F(\phi)$ as a function of $\phi$, called Faraday dispersion function or Faraday spectrum, are a Fourier transform pair. By applying the RM synthesis~\citep{Brentjens2005} technique, $F(\phi)$ can be derived. The limited observation bandwidth with some channels flagged due to Radio Frequency Interference (RFI) introduces a sample function in $\lambda^2$ domain, which results in a spread function, called the RM spread function (RMSF), in the $\phi$ domain. Similarly to the clean procedure in imaging, RM CLEAN~\citep{Heald2009} was developed to estimate the true $F(\phi)$.  

The Faraday dispersion function $F(\phi)$ reflects the emission structure of a polarized source as a function of Faraday depth. In the case of a Faraday-thin~\citep{Brentjens2005} polarized source behind a rotation only magneto-ionic medium, $|F(\phi)|$ will manifest as a single peak at $\phi_{\rm peak}={\rm RM}$ with $|F(\phi_{\rm peak})|=P$, where $P$ is the polarized intensity of the source. The width of the peak is equal to the full width at half maximum (FWHM) of the RMSF. Therefore, the FWHM determines how precisely the RM can be measured. In more complicated cases, such as Faraday-thick components or multiple emission components, $|F(\phi)|$ will show broadened multiple peaks. 

We use \texttt{rmsynth3d} and \texttt{rmclean3d} in the \texttt{RM-Tools} package~\citep{Purcell2020,VanEck2025} to apply RM synthesis and RM CLEAN on both 1-MHz channel and 18-MHz channel images to obtain the Faraday spectra. The FWHM of the RMSF is 59\,rad\,m$^{-2}$ . 
We then use \texttt{rmtools\_peakfitcube} in \texttt{RM-Tools} package to find the fitted $\phi_{\rm peak}$ and $|F(\phi_{\rm peak})|$, corresponding to $\rm RM$ and $P$. The noise in the Faraday cubes has a mean rms $\sigma_P$ of \(\sim\)13.5~$\mu$Jy~beam$^{-1}~$RMSF$^{-1}$.

\subsection{X-ray}

In order to study potential counterparts from X-ray, we make use of archival datasets to highlight the diffuse emission. For a detailed \textit{Chandra}, \textit{XMM-Newton}, and \textit{NuSTAR} study, see \citet{Sander2026}.

\subsubsection{Chandra} 
To compare the radio data with X-ray diffuse emission, we present the Chandra ACIS-S observation of PSR J1357$-$6325 (ObsID: 10880). The observation was taken on August 8, 2009, with a total exposure time of 69.22 ks. The global properties of the X-ray emission were previously presented by \citet{Chang2012}. We reprocessed the data using \textsc{CIAO} version 4.15 \citep{Fruscione2006} and calibration files (CALDB version 4.10.7), including additional steps to enhance the visibility of diffuse emission.

The \texttt{wavdetect} tool in CIAO utilizes a Mexican-hat wavelet to identify discrete sources, which are then employed for aligning images and subsequently removed to analyze diffuse emissions. A diffuse (discrete source-excised) X-ray intensity image was produced with a smoothing optimized for comparison with the radio data following the approach of \citet{Klingler2016}. Note that the final diffuse image has undergone significant processing, and caution is advised when interpreting its scientific properties\footnote{\url{https://cxc.cfa.harvard.edu/ciao/threads/diffuse_emission/}}. This research employs a list of Chandra datasets, obtained by the Chandra X-ray Observatory, contained in~\dataset[DOI: 10.25574/cdc.519]{https://doi.org/10.25574/cdc.519}.

\subsubsection{eROSITA}
We extracted public data from the eRASS:1 release of the SRG/eROSITA All-Sky Survey \citep{Predehl2021,Merloni2024}. The region containing G309.8$-$2.6 is covered by four eROSITA tiles (205153, 204156, 212153, and 211156). These data were processed by the eSASS (eROSITA Standard Analysis Software) pipeline~\citep[][processing number \#020]{Brunner2022}, and have an average angular resolution of 26$\arcsec$. For data analysis, we use the standard tasks available in eSASS. 
\begin{figure*}[hbtp!]
  \centering
  \includegraphics[width=0.9\linewidth]{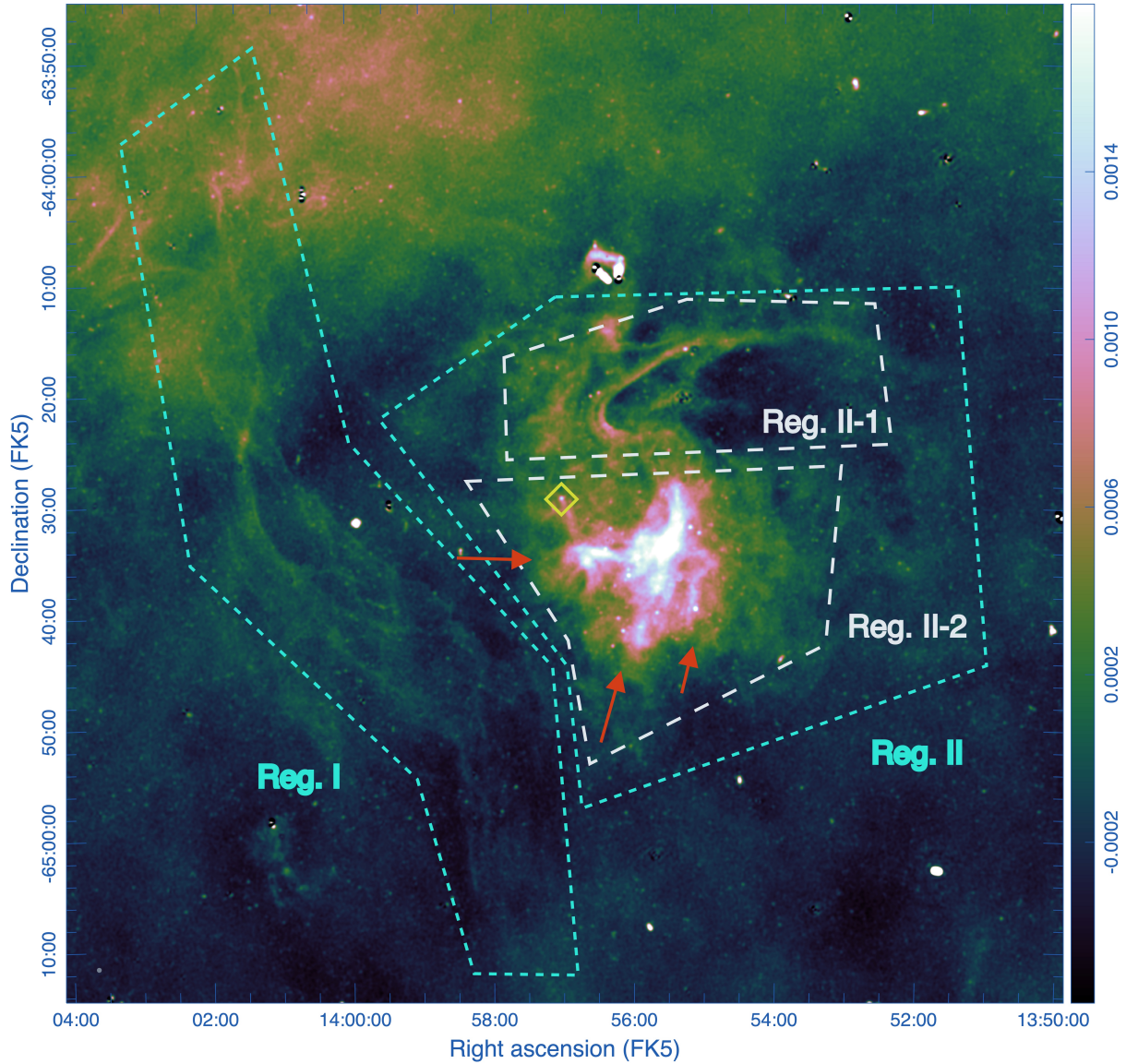}
  \caption{The EMU and STAPS total intensity combined image of the G309.8$-$2.6 at 943 MHz, where the background and point sources have been subtracted. The resolution of this image is 18$\arcsec$. The location of PSR J1357$-$6429 is illustrated by a yellow diamond. Cyan and white polygons mark the primary emission structures with the texts showing the nomenclature, and red arrows indicate the three limbs discussed in Section \ref{sec:overall-radio}.}
  \label{fig:radio-overall}
\end{figure*}

We created intensity maps for the two energy bands 0.5$-$2.3 keV (soft) and 2.3$-$7 keV (hard), used \textit{erbox} to detect the point-like sources, and masked all those sources. The 0.2–0.5 keV band image was excluded here as there was no significant emission. The final image was smoothed to a resolution of 4$\arcmin$ to enhance the visibility of the diffuse emission.
\subsection{Other data}

To further investigate the nature of this object, we incorporated multi-wavelength observations. In radio, we examined maps from the GaLactic and Extragalactic All-sky MWA eXtended (GLEAM-X; \citealt{Ross2024}), the GaLactic and Extragalactic All-sky MWA (GLEAM; \citealt{Hurley-Walker2017}), HI4PI: A Full-Sky neutral hydrogen (H\,\textsc{i}) Survey Based on EBHIS and GASS \citep{Bekhti2016}, the Parkes 2.4 GHz Survey of the Southern Galactic Plane \citep{Duncan1995}, and the Parkes-MIT-NRAO (PMN) survey \citep{Condon1993}. For infrared data, we use all-band observations from the Wide-field Infrared Survey Explorer (\textit{WISE}) \citep{Wright2010}. In the optical, we use maps from the Southern H-Alpha Sky Survey Atlas (SHASSA) \citep{Gaustad2001}, and catalog data from Gaia Data Release 3 (Gaia DR3) \citep{GaiaCollaboration2023}. For comparison in the high-energy regime, we also retrieved the High Energy Stereoscopic System (H.E.S.S.) image of HESS J1356$-$645\footnote{\url{https://www.mpi-hd.mpg.de/hfm/HESS/hgps/}}~\citep{H.E.S.S.Collaboration2018}.

In H$\alpha$, we also retrieved the H$\alpha$ image of the region from the AAO/UKST SuperCOSMOS H$\alpha$ Survey\footnote{\url{http://www-wfau.roe.ac.uk/sss/halpha/hapixel.html}} \citep{Parker2005}. This image is continuum-subtracted, constructed from two deep exposures of 900\,s each. It achieves a sensitivity of $\sim$5~Rayleigh and a spatial resolution of 1$\arcsec$. To cover the full spatial extent of G309.8$-$2.6 and its surroundings, we downloaded 16 adjacent fields, and combined them into a single mosaic using the \textsc{Montage}\footnote{\url{http://montage.ipac.caltech.edu}} software. The resulting mosaic covers an area of $100\arcmin \times 100\arcmin$, with a pixel scale of 0$\farcs$67.

\section{Results}\label{sec:results}

\subsection{Overall radio morphology}\label{sec:overall-radio}

Figure~\ref{fig:radio-overall} presents the combined EMU and STAPS total intensity image of the G309.8$-$2.6 field, as described in Section~\ref{sec:data}. Based on its morphology, we divide the field into several regions, as shown in Figure~\ref{fig:radio-overall}. Here, Region II represents the relic PWN, Region I represents the shell, and these two together represent the SNR. 

At the center of the image is PSR J1357$-$6429 (RA = 13$^{\mathrm h}$57$^{\mathrm m}$02.43$^{\mathrm s}$, Dec = $-$64$\arcdeg$29$\arcmin$30.2$\arcsec$; at the center of the yellow diamond). 
Based on the flux density measurements of $417 \pm 58$\,$\mu$Jy at 1376\,MHz and $309 \pm 86$\,$\mu$Jy at 2496\,MHz by \citet{Kirichenko2015}, we make a power-law fitting and obtain a flux density of $0.56 \pm 0.18$\,mJy at 943\,MHz.
However, the pulsar exhibits an integrated flux density of $1.46 \pm 0.40$\,mJy, significantly higher than the fitted value. This discrepancy is attributed to the blending of the pulsar with its associated compact PWN, as discussed in \citet{Chang2012}. The Stokes $V$ image from EMU (not shown in this work) reveals a circular polarization of $0.16 \pm 0.09$\,mJy, leading to a lower limit of a fractional circular polarization ($|V|/I$) of $\sim11$\%. 
A small fractional circular polarization can also be seen from the profiles by \citet{Kirichenko2015}.

In the vicinity of the pulsar, various filamentary structures and protrusions are observed (see Figure~\ref{fig:pwn}). However, none of these features spatially coincide with the X-ray PWN G309.92$-$2.51 (see Section~\ref{sec:xray} for details). The most prominent radio feature is a bridge-like structure extending southward from the pulsar. Its surface brightness initially increases with distance along the bridge, peaks at about $1\farcm5$ from the pulsar, and then gradually fades as it merges into the broader emission region. This morphology suggests a physical connection between the pulsar and the surrounding extended radio emission.

Beyond the surroundings of the pulsar, the larger-scale diffuse emission exhibits a complex filamentary morphology (i.e., the Salamander; Figure~\ref{fig:radio-overall}, Region II). The main structure traces an S-shaped feature, and its southern arc bifurcates near RA = 13$^{\mathrm h}$55$^{\mathrm m}$00$^{\mathrm s}$, Dec = $-$64$\arcdeg$35$\arcmin$00$\arcsec$ (J2000), forming three limbs extending toward the east, south, and southwest then turning to southeast (as indicated by the red arrows in Figure \ref{fig:radio-overall}). This structure spans an angular size of $\sim$35$\arcmin$, and was identified by \citet{H.E.S.S.Collaboration2011} as the main component of the relic PWN. We measure the flux density of this relic PWN using both the combined image and the ASKAP-only image, after independently subtracting the background and compact sources. The resulting flux density ranges from 2.0 to 3.1 Jy, which is consistent with the previous estimate of $2.8 \pm 0.4$ Jy by \citet{Ball2025}. Note that even if the missing flux density could be compensated with the STAPS data, the flux density could be underestimated due to background subtraction. In addition, several compact sources are entangled with the bulk emission. These could be unrelated background objects or compact knots intrinsic to the relic PWN. Given the complexity of the background, these sources cannot be reliably disentangled in the present data.

In the eastern part of Figure \ref{fig:radio-overall} (Region I), traces of thin shell-like emission can be seen. This structure is elongated northeast-southwest with an extension of $\sim$$1\arcdeg$. Its morphology is suggestive of a highly compressed shell, particularly in later stages of its evolution; we call it ``shell'' in the following analysis. The location of the partial shell's geometric center, slightly west of PSR J1357$-$6429, suggests that the shell is the pulsar's host remnant and that the pulsar is moving eastward. The peak intensity of the shell-like emission reaches an intensity of 100–200\,$\mu$Jy\,beam$^{-1}$, corresponding to a signal-to-noise ratio of up to $\sim$10. 
\begin{figure}[htp]
  \includegraphics[width=\linewidth,trim=0.1cm 1cm 0.1cm 0cm, clip] {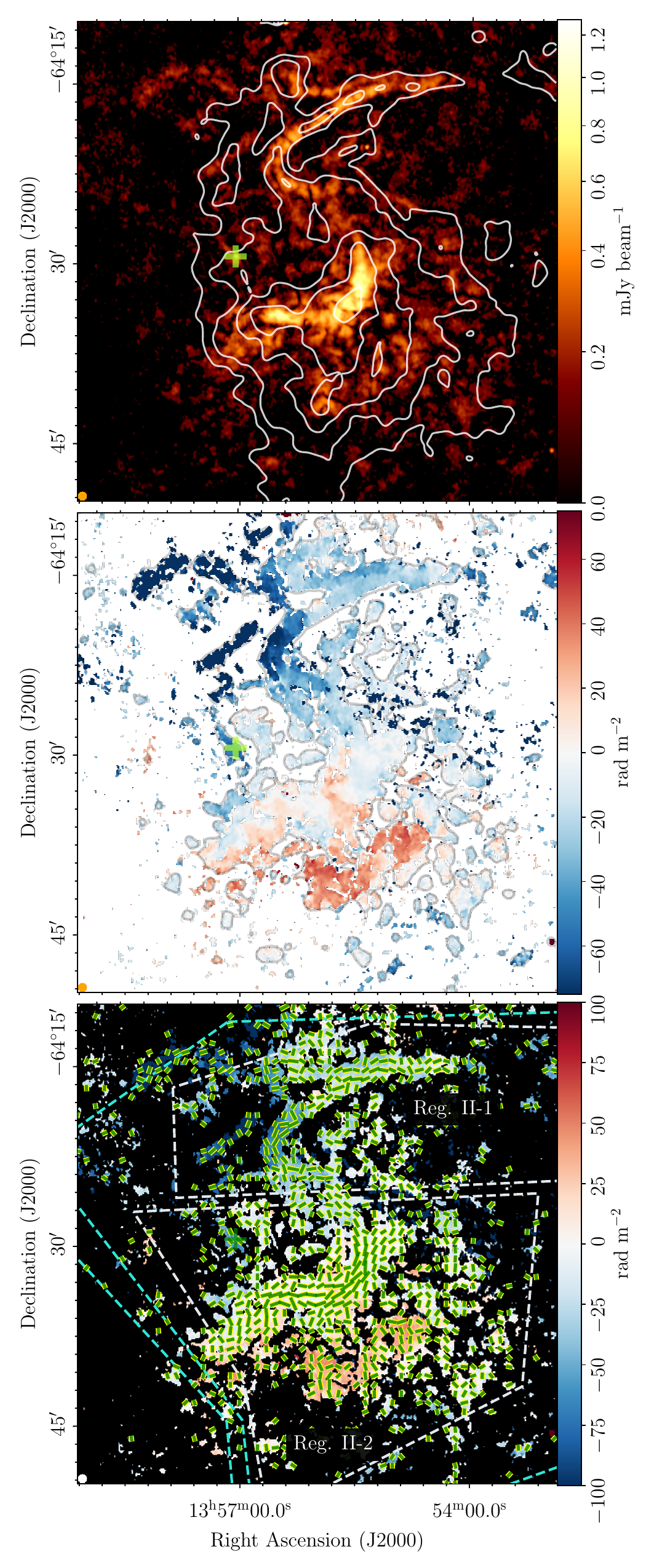}
  \caption{Upper panel: Peak linearly polarized intensity map at 943\,MHz in square root color scale, with white contours for Stokes~$I$ at levels of 5$\sigma_I$, 10$\sigma_I$, and 15$\sigma_I$. Middle panel: RM map, where pixels with $P < 8\sigma_P$ are masked. The contours outline the boundary of $P$. Lower panel: Same as the middle panel, but overlaid with bars showing the orientation of the transverse magnetic field, corrected for Faraday rotation. The bar length is proportional to the square root of the magnitude of $P$. Dashed lines mark the regions defined in Figure~\ref{fig:radio-overall}. 
In all panels, a cross marks the position of PSR~J1357$-$6429.}
  \label{fig:relic-pi}
\end{figure}

Determining the spectral properties of G309.8$-$2.6 is challenging due to frequency-dependent missing short spacings and the limited total bandwidth, and existing higher-frequency single-dish data from Parkes are severely contaminated by strong unrelated sources. Nevertheless, we use sub-band images with a bandwidth of 72\,MHz to investigate the spectral indices of Region I and Region II separately. For Region II (relic PWN), we obtain a spectral index of $\alpha$$\sim$$-0.5\pm0.2$, in contrast to the nearly flat spectrum ($\alpha \sim 0$) reported by \citet{H.E.S.S.Collaboration2011}. The difference may be caused by the regions used to derive the integrated flux density. However, this measurement should be treated with caution, given the limitations of the data. For Region I (the shell), its faintness and the complexity of the surrounding region prevent a reliable measurement of the spectral index or even the integrated flux.

Overall, the G309.8$-$2.6 field presents a complex radio morphology, comprising a PWN powered by the central pulsar PSR J1357$-$6429, an extended relic PWN with tangled filaments, and a faint partial shell structure. The shell lies on the eastern side of the pulsar, while the relic PWN is located to the west.

\subsection{Polarization properties}

Maps of peak Faraday dispersion function, $|F(\phi_{\rm peak})|$ and the peak Faraday depth, $\phi_{\rm peak}$, at 943 MHz, overlaid with the orientation of transverse magnetic field corrected for Faraday rotation, are shown in Figure \ref{fig:relic-pi}. We refer to $|F(\phi_{\rm peak})|$ and $\phi_{\rm peak}$ as $P$ and $\rm RM$ hereafter, respectively. The white contours represent Stokes $I$, starting at 5$\sigma$. To reveal a more coherent structure, we applied a conservative mask by excluding pixels where $P < 8\sigma_P$.

G309.8$-$2.6 exhibits significant linear polarization, with typical $P$ values $\sim$140\,$\mu$Jy\,beam$^{-1}$. The S-shaped structure seen in total intensity is even more prominent in polarization, with the magnetic field generally aligned tangentially along the arc (see Figure \ref{fig:relic-pi}). We adopt the name ``Salamander'' for the main feature, which extends from tail to head in a north-to-east orientation. This alignment suggests a well-ordered magnetic field, likely shaped by particle acceleration or compression processes, such as those found in relic PWNe, reverse shocks, or aged SNR shells \citep[e.g.,][]{Kothes2006}. In addition, a faint shell-like feature is observed starting at RA = 13$^{\mathrm h}$55$^{\mathrm m}$30$^{\mathrm s}$, Dec = $-64\arcdeg23\arcmin30\arcsec$ (J2000), which appears to connect the northern arc and the outer boundary of the lower part and is also marked as Segment 2 in Figure \ref{fig:I-PI}.

In addition to the main S-shaped feature, faint polarized emission is seen surrounding it, as shown in Figure \ref{fig:relic-pi}. Although this diffuse emission shows partial spatial coherence, fluctuations in $P$ are observed near the peak of the relic PWN, which is consistent with expectations for compressed or toroidal magnetic fields. Moreover, several regions located outside the main structure exhibit polarized emission without corresponding features in total intensity. These detections are likely due to smooth large-scale emission in the Galactic foreground magneto-ionic medium, which is mostly filtered out in the ASKAP image cube due to missing short spacing emission. Such anti-correlations between total and polarized intensity are commonly observed in polarization surveys \citep[e.g.][]{Landecker2010,Gaensler2025,Sun2025}.

The fractional polarization, $p$, is estimated as:
\begin{equation}
  p = \frac{P}{I_{\mathrm{src}}},
\end{equation}
where $I_{\mathrm{src}}$ represents the background-subtracted Stokes $I$ emission. We define $I_{\mathrm{src}} = I_{\mathrm{combined}} - I_{\mathrm{bkg}}$, where $I_{\mathrm{combined}}$ is the total intensity from the combined ASKAP and Parkes data, and $I_{\mathrm{bkg}}$ is the fitted background emission. As detailed in Section~\ref{sec:data}, the background-subtracted combined image (Figure \ref{fig:radio-overall}) and the ASKAP-only image ($I_{\mathrm{ASKAP}}$, Figure \ref{fig:SD-Int} right panel) exhibit similar flux density levels. This similarity suggests that estimates of fractional polarization derived from $P/(I_{\rm combined} - I_{\rm bkg,\,combined})$ and $P/ (I_{\rm ASKAP}-I_{\rm bkg,\,ASKAP})$ are similar, where the background of the ASKAP-only image is nearly zero.

Both estimation methods indicate $p$ ranging from 5\% to 15\% in the southern arc (Region II-2). However, the northern arc (Region II-1) displays unexpectedly high values, frequently exceeding 70\%. For synchrotron radiation in a uniform magnetic field, the maximum linear polarization is $p=(s+1)/(s+7/3)$, where $s$ is the electron energy index in the distribution $N(E) \propto E^{-s}$. For $s=2$ corresponding to $\alpha=-0.5$, this gives $p \approx 69\%$~\citep{Reynolds2012}. Therefore, the observed values of $p$ are unphysical for synchrotron radiation. This overestimation of $p$ could come from an underestimated total intensity ($I_{\rm src}$) or inaccurate background subtraction. Furthermore, the absence of short-spacing information in the Stokes $Q$ and $U$ maps introduces a significant source of uncertainty, particularly since the spatial distributions of $P$ and $I_{\mathrm{src}}$ show distinct physical structures that may be affected by missing short baselines in different ways~\citep{Ordog2025}.

The RM distribution is generally smooth but shows a clear structure. As shown in Figure~\ref{fig:relic-pi} middle panel, a gradient and reversal in the sign of RM are observed across the source from northeast to southwest, ranging from about $-200$ to $+100$\,rad\,m$^{-2}$. The northern side has mostly negative RM values, while the southern side shows positive values. In particular, centered on the Stokes $I$ peak in Region II-2 (RA = 13$^{\mathrm h}$55$^{\mathrm m}$00$^{\mathrm s}$, Dec = $-$64$\arcdeg$35$\arcmin$00$\arcsec$, J2000), the RM distribution exhibits complex localized variations of $\sim$30\,rad\,m$^{-2}$, showing negative values to the northeast and mostly positive or near-zero values to the southwest. 

\begin{figure}[!htbp]
    \centering
    \includegraphics[width=\linewidth]{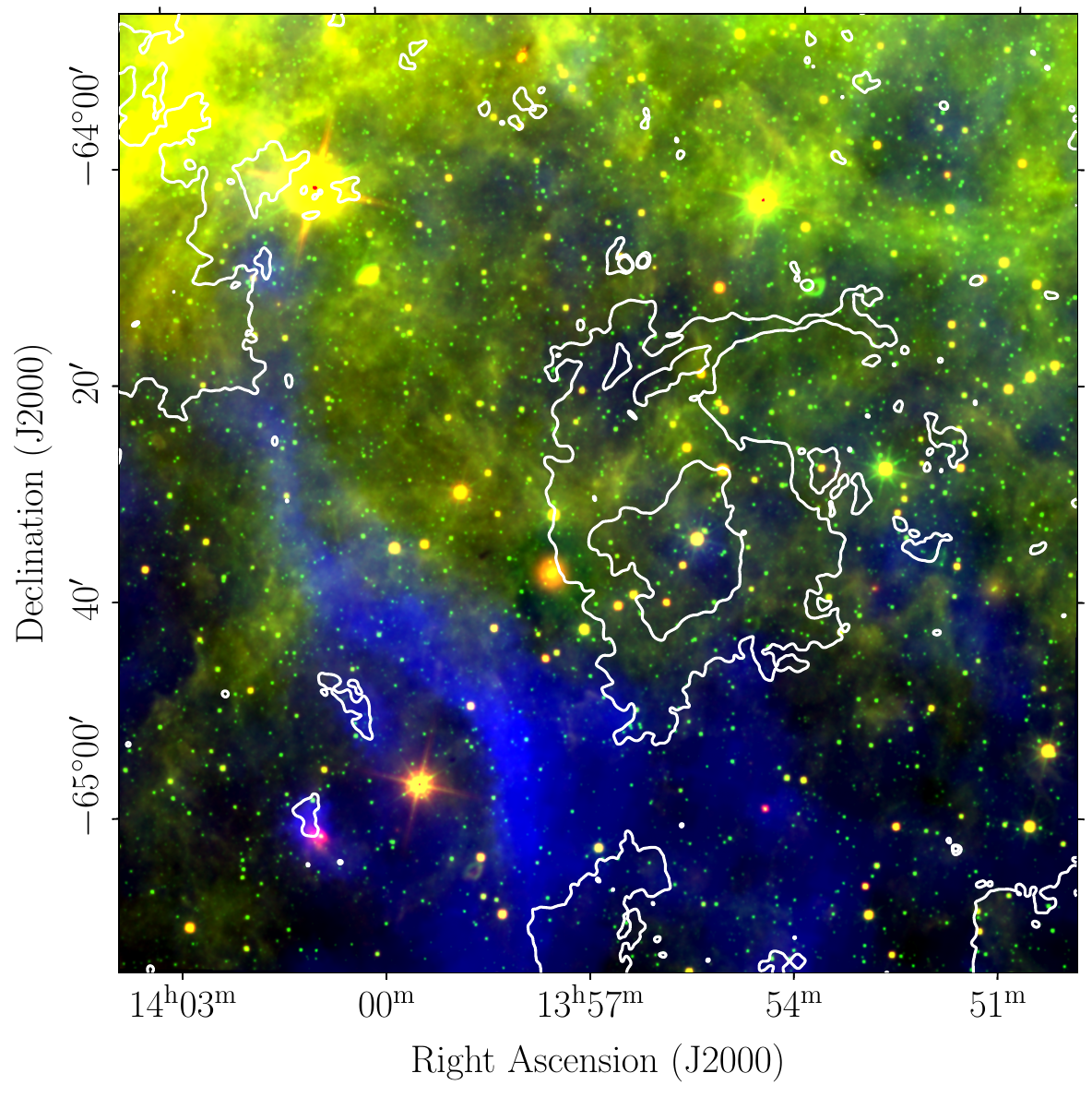}
    \caption{Image of the G309.8$-$2.6 and its surrounding region, with white contours for ASKAP total intensity. \textit{WISE} 22\,$\mu$m emission is shown in red, \textit{WISE} 12\,$\mu$m emission in green, and SHASSA smoothed \protect$\mathrm{H}\alpha$ image in blue.}
    \label{fig:infrared-ha}
\end{figure}
\begin{figure*}[!htbp]
  \centering
  \includegraphics[width=0.95\linewidth]{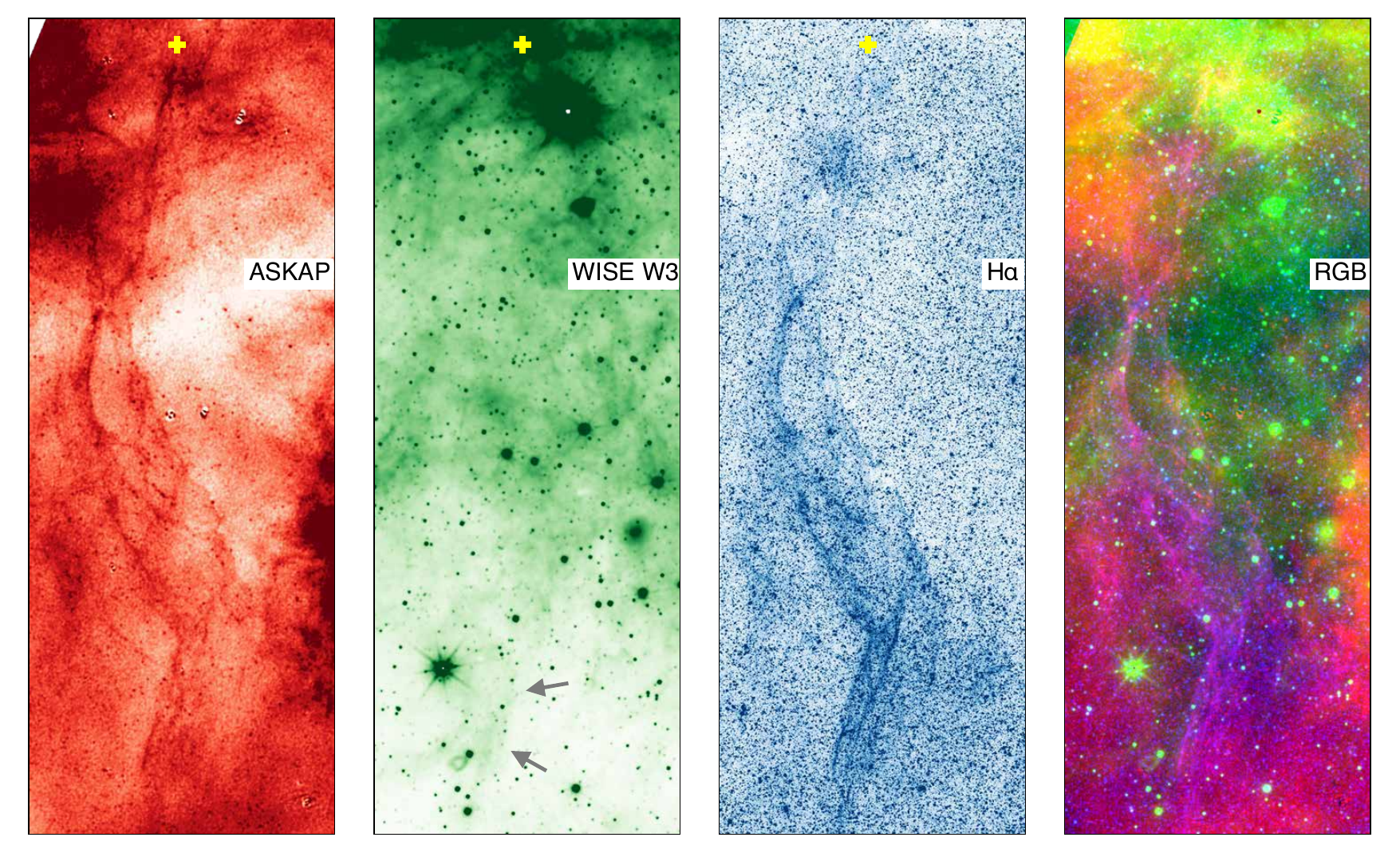}
  \caption{Multi-wavelength view of the shell structure in G309.8$-$2.6. From left to right: ASKAP image at 943 MHz, \textit{WISE} 12 $\mu$m emission, SuperCOSMOS $\mathrm{H}\alpha$ emission, RGB image combining radio (red), infrared (green), and $\mathrm{H}\alpha~$ (blue). The position of PSR J1401$-$6357 is marked as a yellow cross at the first three panel. The grey arrows mark the feature discussed in Section~\ref{sec:possible-shell}.}

  \label{fig:shell}
\end{figure*}

The pulsar PSR J1357$-$6429 is detected in both total and polarized intensity as a compact source (see Figure \ref{fig:pwn}). The measured RM at 943\,MHz is $\rm -49 \pm 2$\,rad\,m$^{-2}$. Near the pulsar, polarized emission is observed along a narrow tail-like feature. Interestingly, this feature extends in a direction opposite to the total intensity tail.

\subsection{Infrared}
From \textit{WISE} 12\,$\mu$m and 24\,$\mu$m data (Figure \ref{fig:infrared-ha}), faint emission is detected between the two white contours, roughly outlining the radio relic PWN and partially overlapping with the X-ray emission. These features may trace shock-heated dust associated with the relic PWN. 
No known H\,\textsc{ii} regions in the \textit{WISE} H\,\textsc{ii} region catalog coincide with the remnant. Furthermore, there is no IR counterpart of the shell that supports the non-thermal interpretation.

\subsection{H$\alpha$}\label{sec:possible-shell}
The higher-resolution SuperCOSMOS H$\alpha$ survey image~\citep{Parker2005} reveals faint filamentary features along the southeastern edge of the radio shell. A rotated cutout of this region is shown in Figure~\ref{fig:shell}, 
alongside the ASKAP 943\,MHz and \textit{WISE} 12\,$\mu$m images. 

The H$\alpha$ filaments closely follow the ASKAP radio contours but have no obvious counterpart in the 12\,$\mu$m infrared image. The lack of infrared emission may imply a relatively cold ISM, supporting the interpretation that the SNR is currently in a radiative phase. Notably, the \textit{WISE} 12\,$\mu$m image shows a faint filamentary feature near the southeastern edge of the shell (see Figure~\ref{fig:shell} second panel, indicated by grey arrows), slightly ahead of the radio boundary. The 12\,$\mu$m emission primarily traces stochastically heated small dust grains. It is possible that this emission arises from dust grains heated by the SNR shock front or swept-up ejecta.
\begin{figure*}[!htbp]
  \centering

  \includegraphics[width=\linewidth]{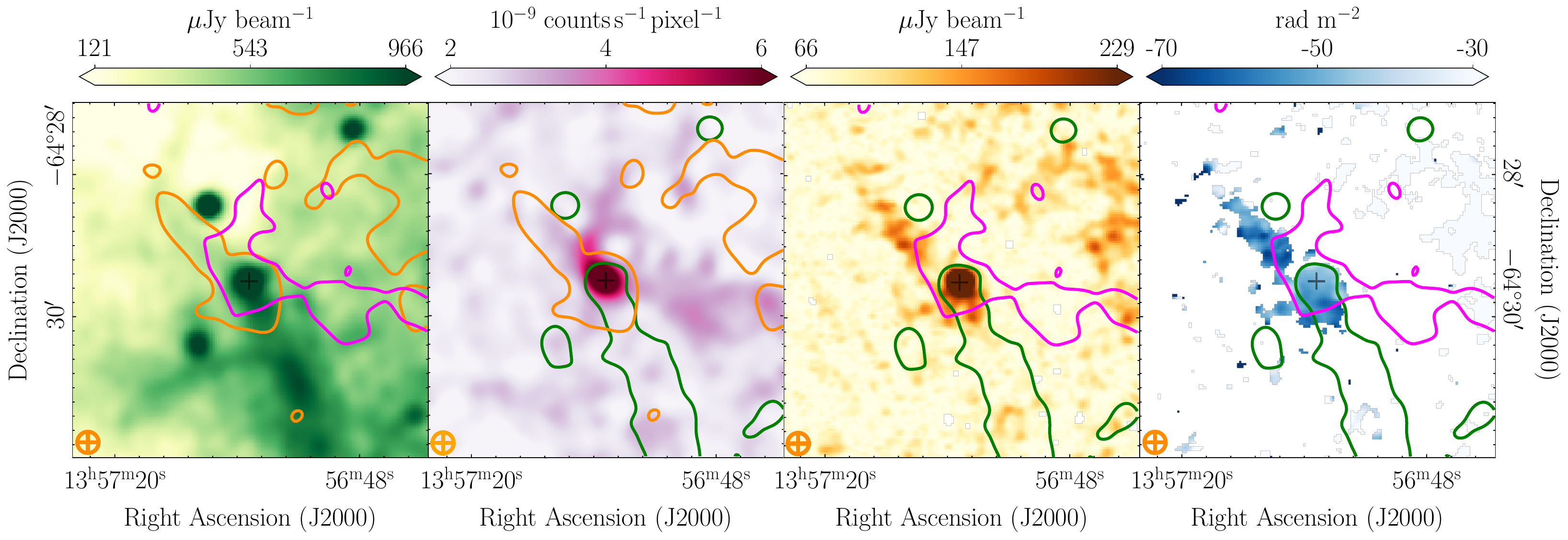}
  \caption{Images of $4\arcmin \times 4\arcmin$ centered on PSR J1357$-$6429. From left to right: ASKAP Stokes $I$ image at 943 MHz, X-ray image in the 0.5--7 keV band from Chandra, $P$ map, and RM map. In all panels, green, magenta, and orange contours represent the Stokes $I$ emission, Chandra diffuse emission, and $P$, respectively. In order to highlight the features, the contours are drawn from slightly smoothed data and the levels are 700 $\mu$Jy beam$^{-1}$, 2.85$\times$10$^{-9}$ counts s$^{-1} $ pixel$^{-1}$, 100 $\mu$Jy beam$^{-1}$.} All images are convolved to a common resolution of 18$\arcsec$, as shown in the lower left corner of each panel. The grey cross marks the position of PSR J1357$-$6429. 
  \label{fig:pwn}
\end{figure*}
A previous study by \citet{Walker2001} reported numerous non-radiative $\mathrm{H}\alpha$ filaments across this region. However, no obvious radio counterparts to these sharp $\mathrm{H}\alpha$ filaments have been detected. \citet{Walker2001} also noted that images obtained \textbf{using the MSSSO Wide Field CCD H$\alpha$ Imaging Survey}~\citep{Buxton1998} reveal emission in both $\mathrm{H}\alpha$ and $[\rm S\,\textsc{II}]$. As the original data are unavailable and the $[\rm S\,\textsc{II}]$/$\mathrm{H}\alpha$ ratio cannot be assessed, it is thus difficult to determine whether the emission arises from shock-heated gas.

\begin{figure}[htbp]
  \includegraphics[width=\linewidth]{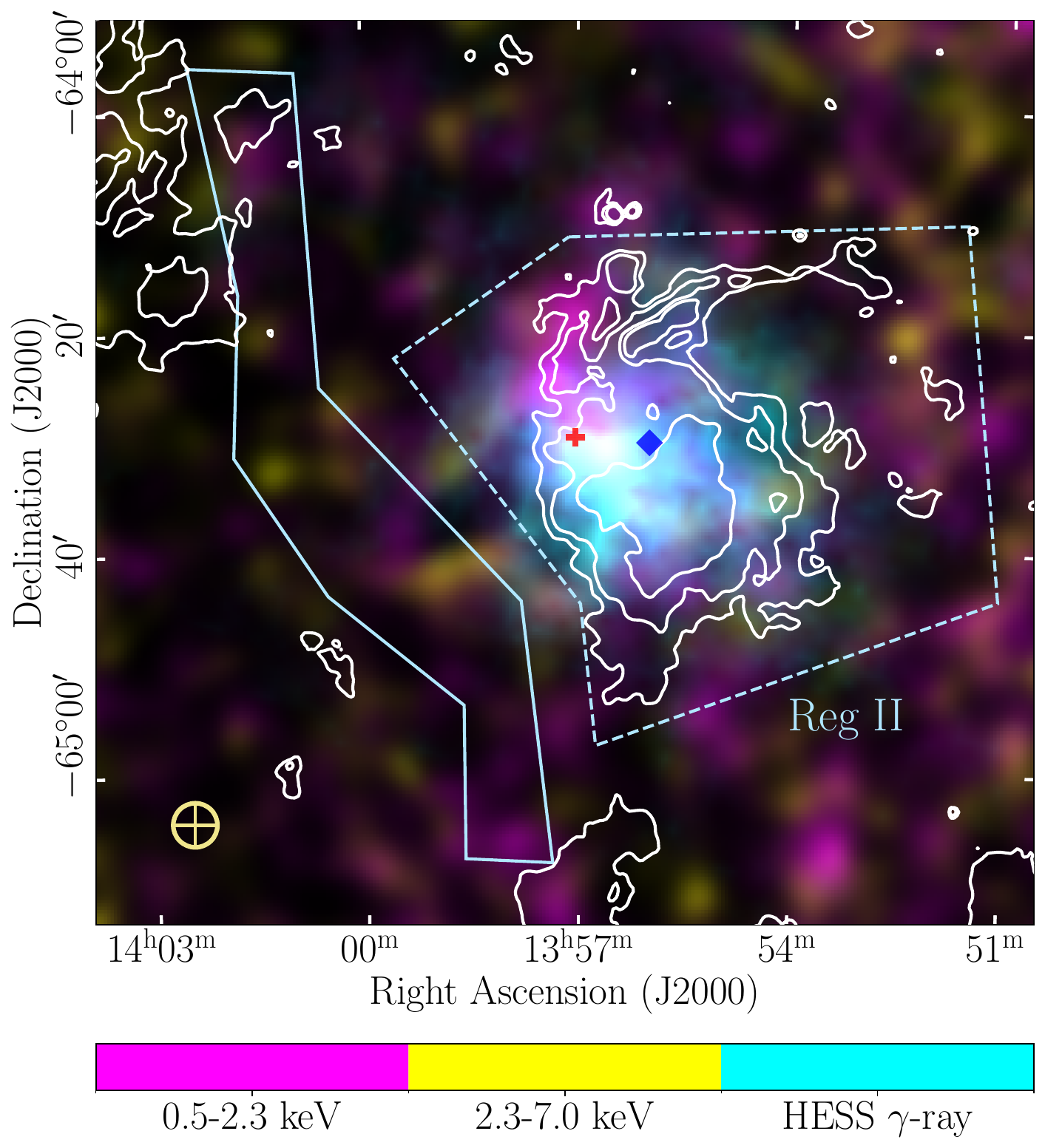}
  \caption{Composite image of X-ray emission from eROSITA and $\gamma$-ray emission from \textit{H.E.S.S.} for the G309.8$-$2.6 field. The eROSITA data are smoothed to a resolution of $4\arcmin$ (shown in the left-bottom corner). The color coding is as follows: 0.2–2.3\,keV in magenta, 2.3–7.0\,keV in yellow, and $\gamma$-ray ($>$0.8 TeV) in cyan. Overlaid white contours are from the 943\,MHz $I$. The skyblue solid polygon shows the boundary of the radio shell, and the skyblue dash polygon indicates the region defined in Figure~\ref{fig:radio-overall}.} The red cross marks the position of PSR J1357$-$6429, and the blue diamond indicates the best-fit centroid of HESS J1356$-$645.
  \label{fig:erosita}
\end{figure}
Additionally, we also notice a nearby pulsar, PSR~J1401$-$6357, located near the terminus of both the radio and $\mathrm{H}\alpha$ shell, as illustrated in Figure~\ref{fig:shell}. This pulsar lies at a distance of $\sim$1.8\,kpc based on its DM of 97.74\,cm$^{-3}$\,pc. The associated H$\alpha$ emission may indicate the proper motion of the pulsar, as fast-moving pulsars can produce a bow shock or trailing tail structure in H$\alpha$~\citep{Brownsberger2014a}. However, the radio filament does not directly connect to the pulsar, no associated X-ray PWN and no proper motion have been reported. These facts argue against a causal relationship between the filament and PSR~J1401$-$6357.

Toward the relic PWN, no large-scale H$\alpha$ emission is detected. However, a faint patch of H$\alpha$ emission can be marginally seen near the western edge of the radio relic. 

\subsection{X-rays}\label{sec:xray}
The \textit{Chandra} diffuse image reveals a compact, tail-like PWN extending in a northeast–southwest direction from PSR J1357$-$6429, as shown in Figure~\ref{fig:pwn}. 
This X-ray PWN is not spatially coincident with the radio total or polarized intensity features. Although faint X-ray emission appears to align with the radio tail, the low signal-to-noise ratio and chip gap between the CCDs make it difficult to draw firm conclusions. As the proper motion of the pulsar remains unmeasured, the origin and nature of this elongated structure are still uncertain. We discuss possible interpretations in Section~\ref{sec:discussion}.

Compared to earlier pointed observations such as those by \textit{XMM-Newton}~\citep{H.E.S.S.Collaboration2011} and \textit{Suzaku}~\citep{Izawa2015}, the eROSITA All-Sky Survey (eRASS) offers a significantly larger field of view (FOV), enabling more complete coverage of extended sources. In comparison with the earlier \textit{ROSAT} All-Sky Survey, eRASS also provides higher sensitivity, improved angular resolution, and a broader energy range. To search for potential X-ray features, we examine false-color composite images constructed from the soft (0.5–2.3~keV) and hard (2.3–7.0~keV) eROSITA bands, the \textit{H.E.S.S.} $\gamma$-ray image, and radio continuum contours over the full field of G309.8$-$2.6, as shown in Figure~\ref{fig:erosita}.

In the soft X-ray band (0.5-2.3~keV), diffuse emission is clearly detected across a substantial portion of the interior of the source. This emission partially overlaps with regions of high radio brightness. The X-ray surface brightness peaks near the geometric center of the SNR, with a local signal-to-noise ratio of $\sim$4. Notably, the $\gamma$-ray emission is spatially close to the extended radio structure. In the hard X-ray band (2.3–7.0~keV), no statistically significant emission is detected. These results are consistent with previous findings. 

We further convolved both soft and hard band images to angular resolutions of $30\arcsec$, $1\arcmin$, and $2\arcmin$ to find the counterparts of the radio features, but no additional structures were revealed. 

\subsection{HI}

We explored the HI4PI data to search for any potential connection between G309.8$-$2.6, referring to the entire region shown in Figure~\ref{fig:radio-overall}, and surrounding HI structures. However, no strongly associated structure was found to correspond to the G309.8$-$2.6. This may be due to the relatively coarse angular resolution of 16$\farcm$2, which limits our ability to resolve HI structures that could be definitively associated with G309.8$-$2.6.

\section{Analysis}\label{sec:analysis}
The morphologies of the shell (Region I) and the extended emission (Region II) shown in Figure \ref{fig:radio-overall} suggest that they share a common origin, where the shell traces the forward shock of the SNR and the extended emission traces the PWN and its interaction with the SNR. We further explore this possibility through multi-wavelength observations below.
\subsection{Distance}

\begin{figure}[htbp!]

  \centering
  \includegraphics[width=\linewidth]{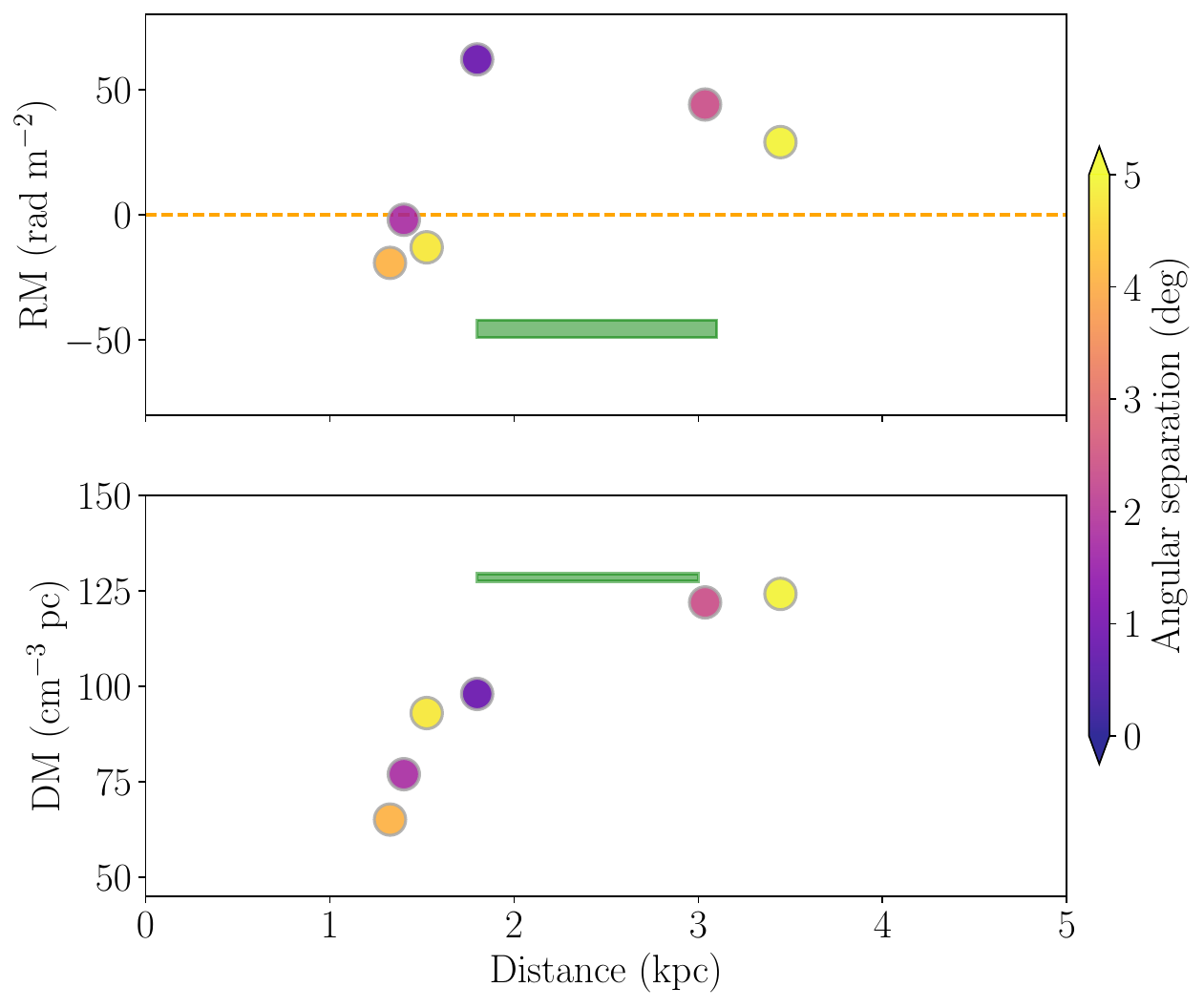}
  \caption{Pulsar RM and DM versus distance. The color represents the angular separation from PSR J1357$-$6429 (PSR J1357). The green rectangles represent the measured RM and DM of PSR J1357 in a distance range of 1.8--3.1 kpc. The orange line indicates RM$=$0.} 
  \label{fig:rm-dm}
\end{figure}

Although the DM-based distance of PSR J1357$-$6429 has been widely used, it has a high uncertainty due to the complex ISM along the LOS. Also, because current electron density models are mostly built using pulsar DMs, using them to estimate distances can lead to a circular argument. The DM-based distance of PSR J1357$-$6429 is estimated to be 3.1\,kpc, based on the YMW16 model~\citep{Yao2017} while the most commonly adopted distance is 2.5\,kpc, which is based on the electron density model from \citet{Cordes2002}. Both distances place the system between the Carina arm and the Crux arm. However, there is some evidence that is consistent with a nearer distance. 
            
We searched for massive stars towards G309.8$-$2.6 from the catalog compiled by \citet{Skiff2014} and \citet{Chen2019}, and derived the parallax distances from GAIA DR3 \citep{GaiaCollaboration2023}. We found a large number of them located at $\sim$1.8\,kpc within the Carina arm. A supernova explosion is most likely to occur in such an environment. Even if we consider the pulsar's motion after the explosion, assuming a typical velocity of 1000~km~s$^{-1}$ and a characteristic age of 7.3~kyr, its projected displacement would be only a few parsecs. Therefore, the PWN and the SNR are expected to be close to the explosion site. This raises the possibility that the relic PWN might be associated with the Carina arm and its massive star population. This may indicate that its actual distance is close to $\sim$1.8 kpc, placing the system within the Carina arm.

We also examine RM and DM versus distance for the pulsars within $5\degr$ of PSR J1357$-$6429 from ATNF pulsar catalog~\citep{Manchester2005} shown in Figure~\ref{fig:rm-dm}. DM increases with distance until about 2~kpc beyond which it is unclear how DM varies with distance, suggesting that PSR J1357$-$6429 could be anywhere between $\sim$2~kpc and $\sim$3~kpc. For RMs, pulsars closer than 1.8 kpc generally exhibit negative values close to zero, whereas those beyond this distance show positive values. It would be more natural to put PSR J1357$-$6429 in the group of pulsars at $\sim$2~kpc, otherwise more reversals of large-scale magnetic field might be needed to interpret its negative RM.

At a distance of 3.1 kpc, the physical size of the relic PWN (Region II) is estimated to be $\sim$31\,$d\rm _{3.1}$ pc, which is significantly larger than the typical size of a relic PWN (e.g., the ``snail'' PWN has a physical size of $\sim$10 pc, \citealt{Ma2016}). At the nearer distance of 1.8 kpc, the physical size is smaller, at $\sim$18\,$d\rm _{1.8}$ pc. Here, $d\rm _{3.1}$ and $d\rm _{1.8}$ are distances in 3.1 kpc and 1.8 kpc, respectively. 

Given this evidence, we prefer a closer distance, but adopt a broad distance range of 1.8–3.1 kpc for our analysis. With this range, the physical size of G309.8$-$2.6 is estimated to be 12$-$18$d_{\rm 1.8}$ pc for the relic PWN (Region II), and 26$-$37$d_{\rm 1.8}$ pc for the eastern shell (Region I) and for the overall system.

\begin{table*}[ht!]
\centering
\caption{Summary of RM surface fitting models}
\label{tab:rm-fit-models}
\begin{tabular}{ccccc}
\hline
\hline
Model & Sample & Number &  Distance Range & Notes \\
\hline
(A) & pulsars in $5\arcdeg$ & 6  & 0–5 kpc & Near-side Galactic contribution \\
(B) & pulsars in $5\arcdeg$ & 20 & 5–10 kpc & Mid-range Galactic contribution \\
(C) & pulsars in $5\arcdeg$ & 11 & $>$10 kpc & Far-side Galactic contribution \\
(D) & pulsars in $5\arcdeg$ & 37 & All distances & Whole Galactic contribution \\
(E) & POSSUM components in $3\arcdeg$ & 417 & -- & Whole Galactic contribution \\

\hline
\end{tabular}
\end{table*}

\begin{figure*}[!htbp]
  \centering
  \includegraphics[width=0.9\linewidth]{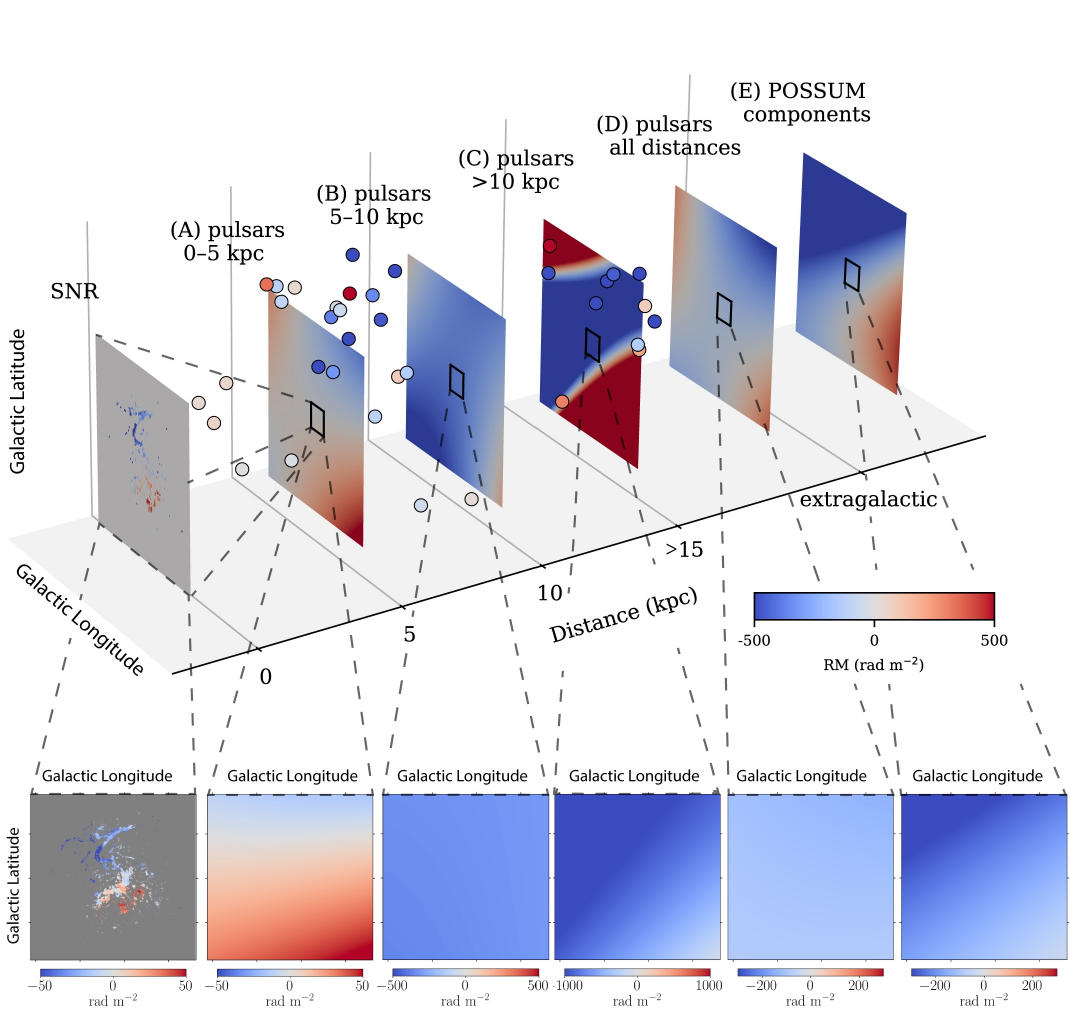}
  \caption{Top panel: 
  The leftmost layer displays the RM map of G309.8$-$2.6, with values ranging from $-20$ to $+20$\,rad~m$^{-2}$. The remaining layers present the fitted quadratic RM surfaces derived from different components, each covering a $6^\circ \times 6^\circ$ region centered on PSR~J1357$-$6429. Annotations above each layer indicate the corresponding component, and the color scale is shared across all layers as shown in the colorbar. The black box in each layer outlines the boundary of G309.8$-$2.6, corresponding to the extent in the leftmost panel. Pulsars are marked as black-edged dots and color-coded by their RM values on the same scale. The $X$ and $Y$ axes represent Galactic longitude and latitude, respectively, while the $Z$ axis indicates distance. 
  Bottom panels: Zoom-in views of the corresponding regions marked by connecting lines in the top panel.}
  \label{fig:rm-fit-dist}
\end{figure*}

\subsection{Polarimetry}

One of the most interesting features of this object is the prominent S-shaped polarization signal and the sign reversal and gradient in RM. As shown in Figure~\ref{fig:relic-pi}, the RM distribution follows an S-shaped morphology with a symmetry about its centre: the northern arc shows predominantly negative RMs while the southern arc displays positive RMs. Since the sign of RM reflects the direction of the electron-density-weighted average line-of-sight magnetic field, such polarization observations of SNRs can provide valuable constraints on the propagation of turbulent magnetic fields in SNR and the isotropic component of turbulence in the plane of the sky.

\subsubsection{Foreground RM}\label{sec:foreground_RM}

The observed RM toward G309.6$-$2.6 arises from Faraday rotation through the magneto-ionic medium, which will always include some contribution from the foreground, likely unrelated foreground interstellar gas~\citep[e.g.,][]{Caswell2004}, or in material swept up by the expanding SNR. To investigate the magnetic field structure intrinsic to the PWN, it is necessary to estimate and, if possible, remove the foreground RM component. 

We use pulsars and polarized extragalactic sources to estimate the large-scale Galactic foreground RM. We use a second-order polynomial surface, $\phi_{\rm{MW}}$, to describe the 2D spatial variation,
\begin{equation}
  \label{eq:quad_surface}
\phi_{\rm{MW}} = A l^2 + B b^2 + C l b + D l + E b + F,
\end{equation}
where $l$ and $b$ are the Galactic longitude and latitude, and $A$ through $F$ are the coefficients. We perform surface fitting using all pulsars within 5$\arcdeg$ of the SNR, taken from the ATNF catalog v2.5.1\footnote{\url{https://www.atnf.csiro.au/research/pulsar/psrcat}}. We also include all polarized components detected in the POSSUM survey within 3$\arcdeg$, to capture the full foreground Galactic contribution \citep[see][for a detailed explanation]{Vanderwoude2024}. In addition, we divide the pulsar sample into three distance bins (0--5 kpc, 5--10 kpc, and $>$10 kpc) to examine the RM variation along the line of sight. In total, we tried five models, which are summarized in Table~\ref{tab:rm-fit-models}, and the corresponding fitted surfaces are shown in Figure~\ref{fig:rm-fit-dist}. 

Model (E) derived from the POSSUM components shows a change of RM sign from negative to positive from the upper-left to the bottom-right in the 6$\arcdeg$$\times$6$\arcdeg$ region. The SNR is located at this boundary of the RM sign transition. The fitted RM surface toward the SNR region (the black square region) has a mean value of $-205 \pm 54$\,rad\,m$^{-2}$. The overall trend in Model (E) is consistent with the RM gradient of the PWN shown in Figure~\ref{fig:relic-pi}. This could indicate that the RM variation that we observe is simply caused by the Galactic foreground or that the SNR is located at the position in the Galaxy where this sign change occurs. However, the POSSUM components are mostly extragalactic and their RMs include the full line-of-sight path to the edge of the Galaxy. Thus, Model (E) represents the total Galactic RM.

Model (A) fitted using the most nearby pulsars with distance 0--5 kpc also shows a change of RM sign, but has a different mean value of $13 \pm 17$\,rad\,m$^{-2}$. In fact, Models (B), (C), and (D) all show different RM patterns. This highlights the complexity of Faraday depth structures along the line of sight and is consistent with multiple sign changes shown in nearby pulsars and predicted in Galactic magnetic field models~\citep[e.g.,][]{Han2018}. The low pulsar density and distance uncertainties further complicate efforts to reliably model the foreground.

Nevertheless, with caution, we adopt the RM surface from Model (A) as an estimate of the foreground contribution and define the intrinsic RM of the PWN as the difference:

\begin{equation}
\mathrm{RM}_{\mathrm{intrinsic}} \approx \mathrm{RM}_{\mathrm{obs}} - \mathrm{RM}_{\mathrm{model}}.
\end{equation}

\begin{figure*}
  \centering
  \includegraphics[width=\linewidth]{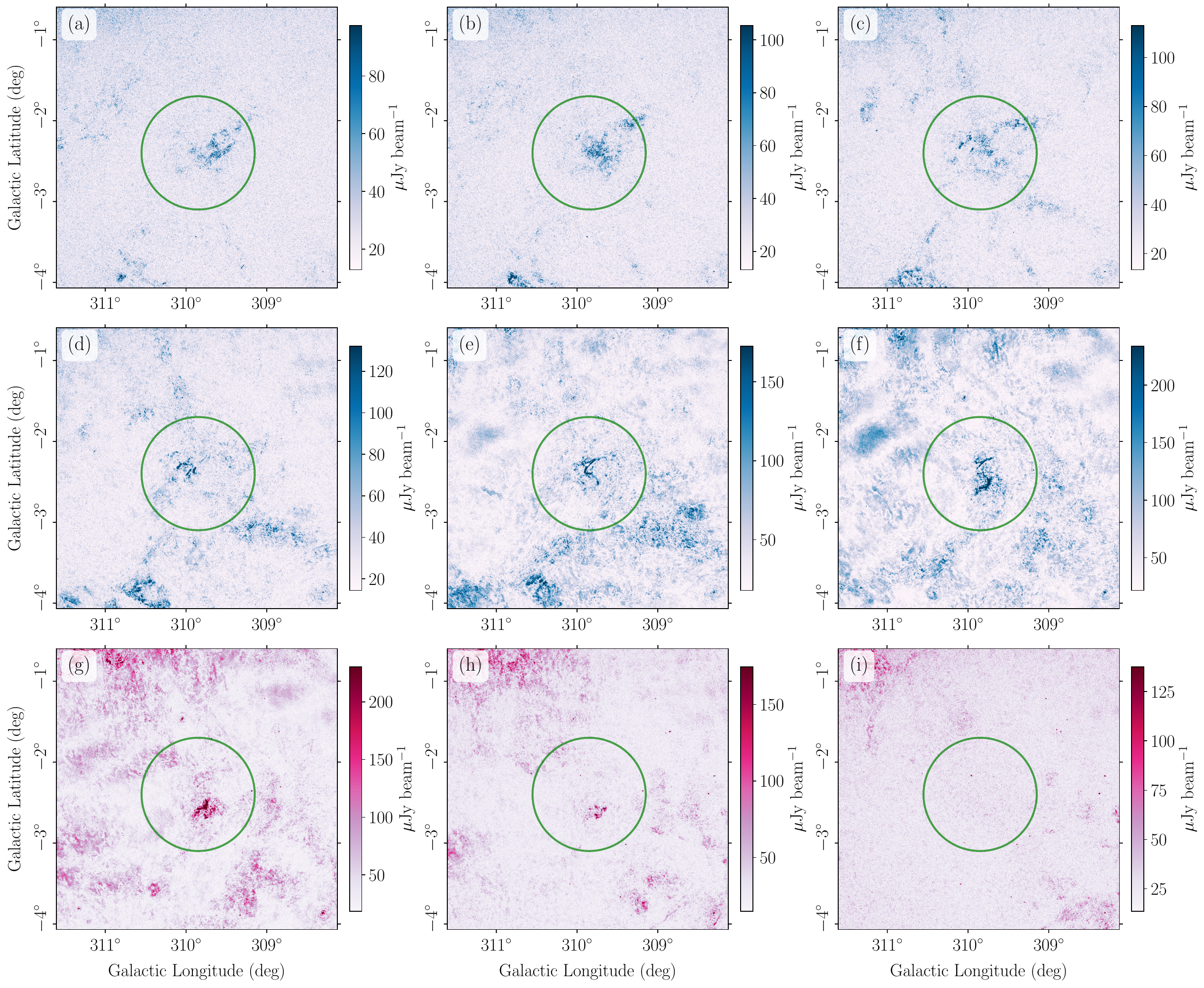}
  \caption{Averaged $|F(\phi)|$ at Faraday depth range of $\rm{-240}$~to~120\,rad\,m$^{-2}$ in a step of 40~rad~m$^{-2}$ to display the large RM distribution separately in Galactic frame. The green circle outlines G309.8$-$2.6. The blue colormap represents negative Faraday depth in panels (a) to (f), and the red colormap represents positive Faraday depth in panels (g) to (i).}
  \label{fig:fdf}
\end{figure*}

\subsubsection{Polarized emission at other Faraday depths}

Besides the main peak in the Faraday spectrum, faint secondary components are visible in several subregions. We extract a $2\arcdeg \times 2 \arcdeg$ region centered on G309.8$-$2.6 from $\phi=-240$ to $+120$~rad~m$^{-2}$, and integrate $|F(\phi)|$ using steps of 40 rad~m$^{-2}$. The average $|F(\phi)|$ for each range of Faraday depth is shown in Figure \ref{fig:fdf}. 

Panels (d) through (h) for the Faraday depth from $-$120 to $+$80 rad m$^{-2}$ show emission that is unambiguously related to the PWN. Panels (a) to (c) show the secondary components in $|F(\phi)|$ from $-$240 to $-$120 rad m$^{-2}$, which are faint and patchy and seem unrelated to the S-shaped PWN based on morphology. The secondary component is well separated from the primary component in Faraday depth. The separation of the order of 100~rad~m$^{-2}$ is larger than the FWHM of the RMSF, indicating that this emission is real.

To produce such a two-component $|F(\phi)|$ structure, at least two synchrotron-emitting regions must exist along the line of sight. It is difficult to conclude the origin of the secondary components. If the secondary components correspond to physically distinct structures within the SNR, they could be indicative of internal shocks, density variations, or interaction zones with the ambient ISM. However, producing the observed large RM separation requires a thermal Faraday screen with large RM between the emission components. Such a Faraday screen is expected to have a very strong magnetic field or a very high electron density. We suggest that the secondary component is more likely from the Galactic ISM.

\section{Discussion}\label{sec:discussion}

Our multiwavelength results and polarization studies reveal a region with significant morphological and environmental complexity. One of the favored explanations is an evolved or relic PWN associated with PSR J1357$-$6429.

As the SNR evolved, the reverse shock propagated inward through the ejecta and interacted with the PWN. This interaction, combined with the pulsar's proper motion and the ambient density gradient, probably disrupted the original, more extended PWN. As a result, a relic PWN was left near the pulsar birth site, while a new, smaller, and possibly more irregular PWN formed around the energetic pulsar currently offset from the centre~\citep{Gaensler2006}.  At the same time, the forward shock of the surrounding associated SNR continued to expand and became distorted by the inhomogeneous environment, potentially causing the partial shell structure observed around the relic PWN. This scenario is consistent with previous studies of the G309.8$-$2.6 and shares key features with other relic PWNe interacting with their host SNRs and surrounding medium, such as Vela~\citep{Slane2018}, ``Boomerang''~\citep[G106.3+2.7,][]{Kothes2006}, CTB 87~\citep{Kothes2020}, the ``Snail'' PWN~\citep{Ma2016}, and the ``Goose'' PWN~\citep{Klingler2022}.

The pulsar PSR~J1357$-$6429 has a characteristic age of $\sim$7.3 kyr and still lies within the remnant, suggesting a middle-aged system in the reverberation phase~\citep{Olmi2023}. If the eastern shell (Region I in Figure \ref{fig:radio-overall}) corresponds to the forward shock, its radius is in the range of $20\arcmin$--$40\arcmin$, namely $R_{\rm s}=$ 10–21\,$d_{\rm 1.8}$ pc. The age can be estimated as $t_{\rm kyr} \approx 0.017(R_{\rm s} / pc)^{5/2} \, (n_0 / E_{51})^{1/2}$ \citep{Sedov1959}, where $n_0$ is the ambient density, $E_{51}$ is the explosion energy in $10^{51}$ erg and $t_{\rm kyr}$ is in kyr. For $E_{51} = 1$ and $n_0 = 0.01$ to 1 cm$^{-3}$,  we derived an age of 3--36\,$d_{\rm 1.8}^{5/2}$\,kyr. This further supports the system to be middle-aged.

\subsection{The Radio Tail Structure}

There is a linear radio structure in $I$ that extends from the X-ray PWN G309.92$-$2.51~(Figure~\ref{fig:pwn}: image in the left panel and green contours in other panels), hereafter referred to as the radio tail structure. The radio tail is not spatially aligned with the X-ray emission.

The nature of the X-ray PWN associated with PSR J1357$-$6429 has previously been discussed by~\citet{Chang2012}, who considered torus-jet and bow-shock interpretations. Diffuse X-ray emission has been found near the pulsar, although its nature remains uncertain. The soft X-ray emission can be fitted with an absorbed power-law model~\citep{Izawa2015}, but the lack of significant spectral variation leaves the origin of the emission open to interpretation~\citep[See][for a detailed discussion]{Sander2026}.

The discontinuity between the emission near the pulsar and that further away along the radio tail suggests that it is not powered by continuous energy injection. Because of the long cooling time of the electrons producing the radio emission at 943\,MHz, which is $\sim10^7$\,yr for a magnetic field of 3.7$\rm \mu G$~\citep[e.g.][]{Liu2023a} and much larger than the age of the pulsar, the radio tail probably traces the past motion of the pulsar PSR J1357$-$6429. However, following the radio tail does not necessarily lead to the geometric center of the host SNR, as shown by idealized symmetric simulations \citep{Kolb2017}. 

The average transverse velocity $V_\perp$ of the pulsar at the head of the radio tail can be estimated using the characteristic age of the pulsar. Assuming an age of $7.3 \text{ kyr}$, this yields $V_\perp=350$--$700\,d_{1.8} \text{ km s}^{-1}$. This range exceeds the mean $V_\perp=$ $307 \pm 47$\,km\,s$^{-1}$ reported by \citet{Hobbs2005}, but remains within the upper limit on the proper motion of $V_\perp=$ 1200 $\text{km s}^{-1}$ derived by \citet{Kirichenko2015}.

However, no bow-shock structure is detected in H$\alpha$ emission although the estimated velocity is very high. Combining with the morphology connected with the relic PWN, it suggests that the pulsar remains embedded within the expanding SNR shell, where the ambient conditions are not yet suitable for a bow-shock to form.

In other cases of escaping pulsars (particularly those exhibiting a bow-shock structure), the radio tail often shows strong linear polarization, typically with tangential or toroidal magnetic field orientations~\citep[e.g.,][]{Ng2010}. However, in our case, the $P$ map in Figure~\ref{fig:pwn} shows no significant polarized emission along the radio tail. This could be caused by depolarization from high RM values or turbulence in the intervening medium. A similar effect has been observed in the ``Goose'' relic PWN, where polarization is seen at 3 and 6 cm with high RM but is absent in lower frequency ASKAP data~\citep{Klingler2022}. 

The linear radio structure may also be interpreted as Rayleigh–Taylor (RT) instabilities that arise during the crushing process as the relativistic gas in the nebula expands into the slow-moving denser ejecta~\citep{Blondin2001}. Synchrotron emission is enhanced along the filaments due to compression of the magnetic field~\citep{Slane2008}. The alignment of the magnetic field with the fainter extension of the tail toward the northwest (Figure \ref{fig:relic-pi}) supports this interpretation. The linear morphology in the synchrotron emission may thus reflect the imprint of RT instabilities as the relic PWN is compressed and reshaped by the reverse shock. These instabilities can also lead to mixing between hot thermal gas and relativistic fluid, further complicating the observed structure~\citep{Blondin2001}.

Toward the pulsar, the RM measured in our ASKAP 943\,MHz data is $-49 \pm 2$\,rad\,m$^{-2}$, consistent with previous measurements at higher frequencies: $-47\pm 2$\,rad\,m$^{-2}$ at 1.369\,GHz~\citep{Johnston2006} and $-43 \pm 1$\,rad\,m$^{-2}$ from 1.376 to 2.358\,GHz~\citep{Kirichenko2015}. Both the ASKAP and the pulsar RM measurements consist of the entire foreground RM and a fraction of the PWN RM. For the ASKAP measurement, the fraction depends on how the thermal and nonthermal electrons are mixed inside the PWN; for the pulsar measurement, the fraction  depends on where the pulsar is located in the PWN. Using the foreground model in Section~\ref{sec:foreground_RM}, the RM of the PWN can be estimated to be $\sim$$-$120~rad~m$^{-2}$. Here, the fraction of 0.5 was used, which means the thermal and non-thermal electrons are uniformly mixed or the pulsar is located at the center of the PWN. However, it is impossible to estimate the magnetic field strength because the thermal electron density and line-of-sight path length are unknown. 

An alternative interpretation of this radio tail is that it might represent part of a jet. The jet extends toward the northeast, where only polarized emission is observed (Figure \ref{fig:pwn}). It is intriguing that the jet is curved in the opposite direction to the curved X-ray emission. Further investigation is needed for this interpretation.

\subsection{The Relic PWN}

The S-shaped structure in Region II (Figure \ref{fig:radio-overall}) exhibits strong linear polarization and clear RM gradient. If these features are intrinsic to the SNR, they may reflect the impact of a reverse shock on the PWN and its magnetic field.

Similar S-shaped structures, where magnetic field lines follow tangential arcs, have also been observed in other young or middle-aged PWNe, particularly in SNR G357.1$-$0.2~\citep{Gray1994,Gray1996,Cotton2022}. 
The highly polarized PWN G328.4+0.2 may also represent a system currently undergoing reverse-shock interaction~\citep{Gaensler2000}. 

\begin{figure}[htbp!]
  \centering
  \includegraphics[width=1\linewidth, trim=0cm 1cm 0.1cm 0cm, clip]{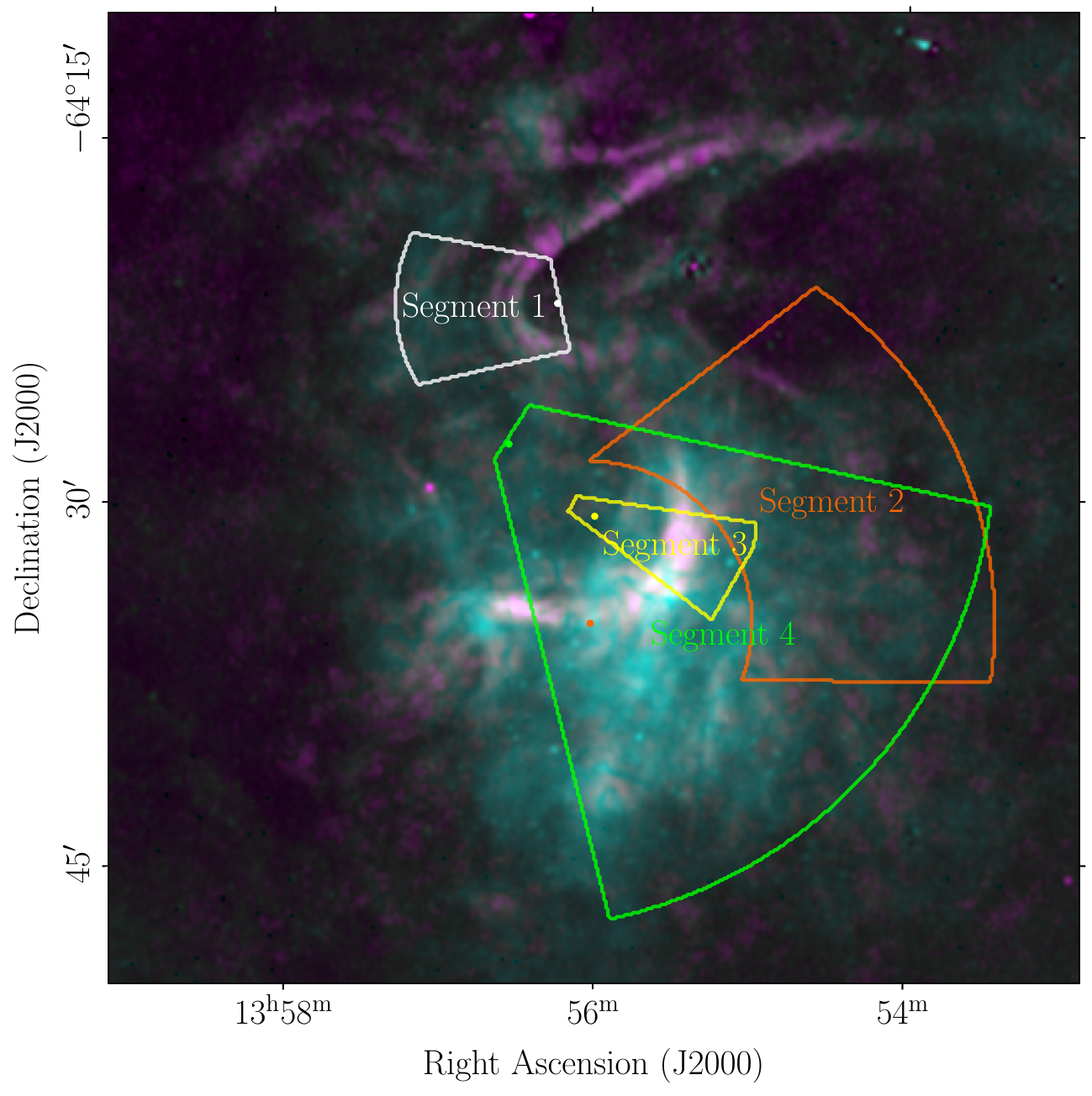}
  \caption{Two-color composite image combining the Stokes $I$ point-source-subtracted image (cyan) and the $P$ image (magenta). The white, orange, yellow, and green fan-shaped regions denote the Segments used for extracting radial profiles shown in Figure~\ref{fig:profile-multi}. The corresponding dots indicate the centers of each Section. }
  \label{fig:I-PI}
\end{figure}

\begin{figure*}[htbp!]
  \includegraphics[width=1\linewidth]{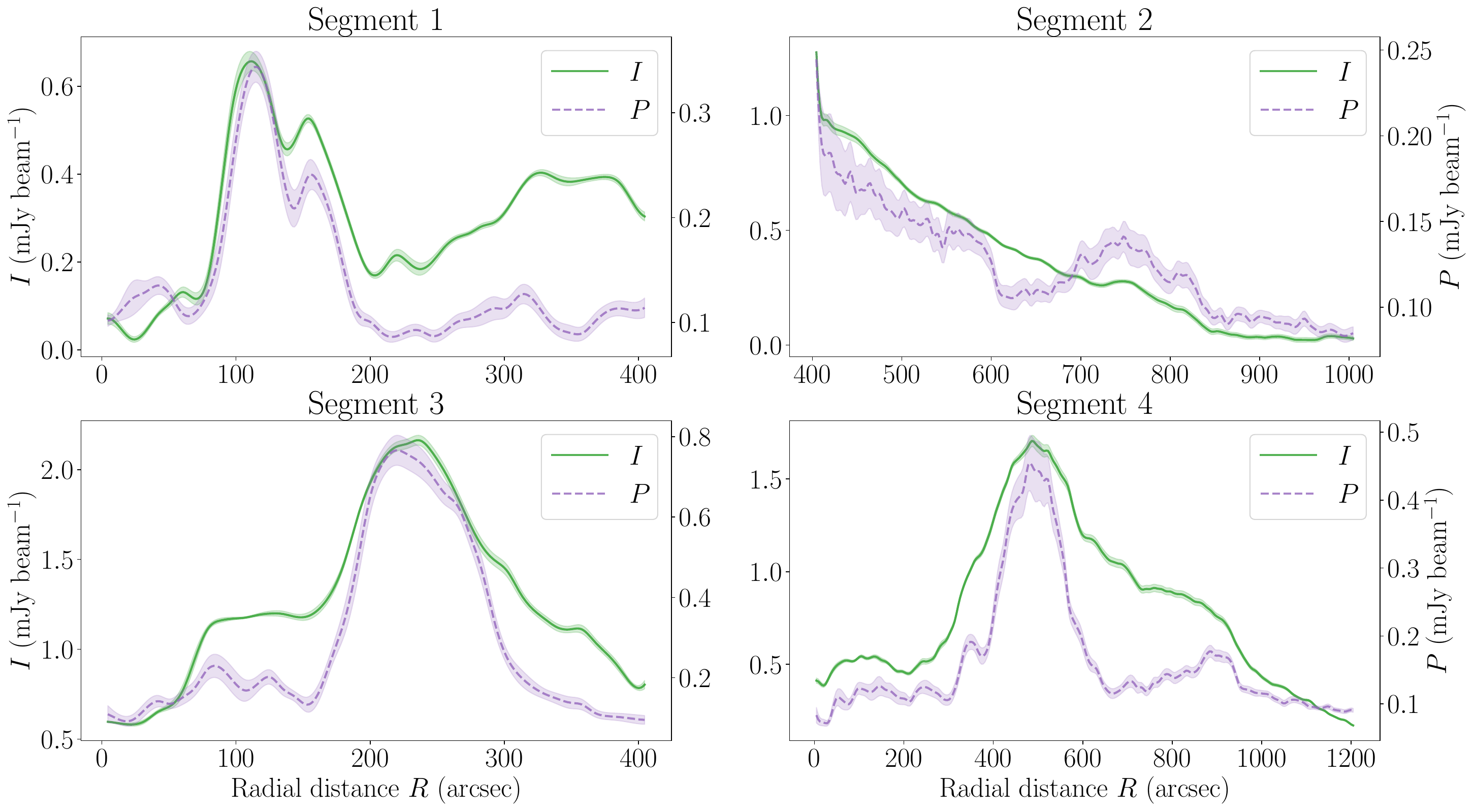}
  \caption{Profiles of $I$ and $P$ versus radius $R$ from the geometric center outward along the radial direction for the four Segments outlined in Figure~\ref{fig:I-PI}.} 
  \label{fig:profile-multi}
\end{figure*}

We combined Stokes $I$ and $P$ images (Figure~\ref{fig:I-PI}) and defined four segments for profile extraction: Segment 1 for the northern arc (Region II-1 in Figure~\ref{fig:radio-overall}) and Segments 2-4 for the southern part (Region II-2 in Figure~\ref{fig:radio-overall}). For each segment, we constructed a set of concentric annuli approximately aligned with the observed arc structure, with the common geometric center marked by a dot in Figure~\ref{fig:I-PI}. Each segment is defined by the intersection between an annulus and the polygon that encloses the area of interest. Profiles of $I$ and $P$  versus radius $R$ were then extracted from the geometric center outward along the radial direction for the four segments, shown in Figure~\ref{fig:profile-multi}.

\subsubsection{Region II-1: the northern part}
In Segment 1, a double-peaked structure is evident in the Stokes $I$ and $P$ profiles~(Figure~\ref{fig:profile-multi}), with the primary peak at $R\approx110\arcsec$ and the secondary peak at $R\approx160\arcsec$, possibly marking the compressed boundaries of the PWN. There is also a weak excess of $P$ beyond these two peaks, which corresponds to an excess of $I$ over a much broader radius range. 
The double-peaked structure possibly traces the reverse shock and moves toward the center of the arc. The sharp inner edge is an unusual feature and is not consistent with the typical morphology expected from the outer shell of an SNR, as can be seen in Figures~\ref{fig:I-PI} and \ref{fig:profile-multi} (Segment 1). The fractional polarization is higher because the turbulent magnetic field is stronger for the already shocked material. The reverse shock heats the gas, which explains the soft X-ray emission outside the two peaks, as can be seen in Figure~\ref{fig:erosita}. 

\subsubsection{Region II-2: the southern part}
Region II-2 in Figure \ref{fig:radio-overall} contains both the southern arc of the S-shaped structure and diffuse emission. Segment 3 in Figure \ref{fig:I-PI} is a zoom-in of part of the arc, and the corresponding radial profiles in Figure \ref{fig:profile-multi} show strong $I$ and $P$ and high relative fractional polarization. Segment 2 in Figure \ref{fig:I-PI} is farther out, and the profiles show a broad weak peak in $P$. Segment 4 contains most of the Region II-2. In addition to showing similar characteristics of Segments 2 and 3, the profiles show an enhancement of $I$ that covers a wide range extending from the bright arc of the PWN. 

The enhancement in $P$ shown in both Segment 3 ($R\sim$200$\arcsec$) and Segment 4 ($R\sim$500$\arcsec$) marks the inner boundary of the relic PWN and indicates a tangential magnetic field, as expected for a relic PWN compressed by reverse shocks from the SNR~\citep{Blondin2001}. The polarization peak in Segment 2 ($R\sim$750$\arcsec$) and Segment 4 ($R\sim$900$\arcsec$) may reflect the forward shock of the SNR. The Region II-2 thus contains both forward shock and reverse shock, which accelerate electrons to produce radio emissions. 

The fractional polarization in the southwestern part of Region II-2 is relatively low, as can be seen in Segment 4 ($R = 500\arcsec-900 \arcsec$, Figure \ref{fig:profile-multi}), implying depolarization. The depolarization might arise from the enhanced turbulent magnetic field due to the interactions between shocks and nearby interstellar clouds~\citep{Jun1999}. The presence of weak $\mathrm{H}\alpha$ emission (Figure \ref{fig:infrared-ha}) supports the existence of such a cloud.

\subsection{Plausible scenarios leading to the observed properties}

\begin{figure*}[t]
  \centering  
  \includegraphics[width=\linewidth]{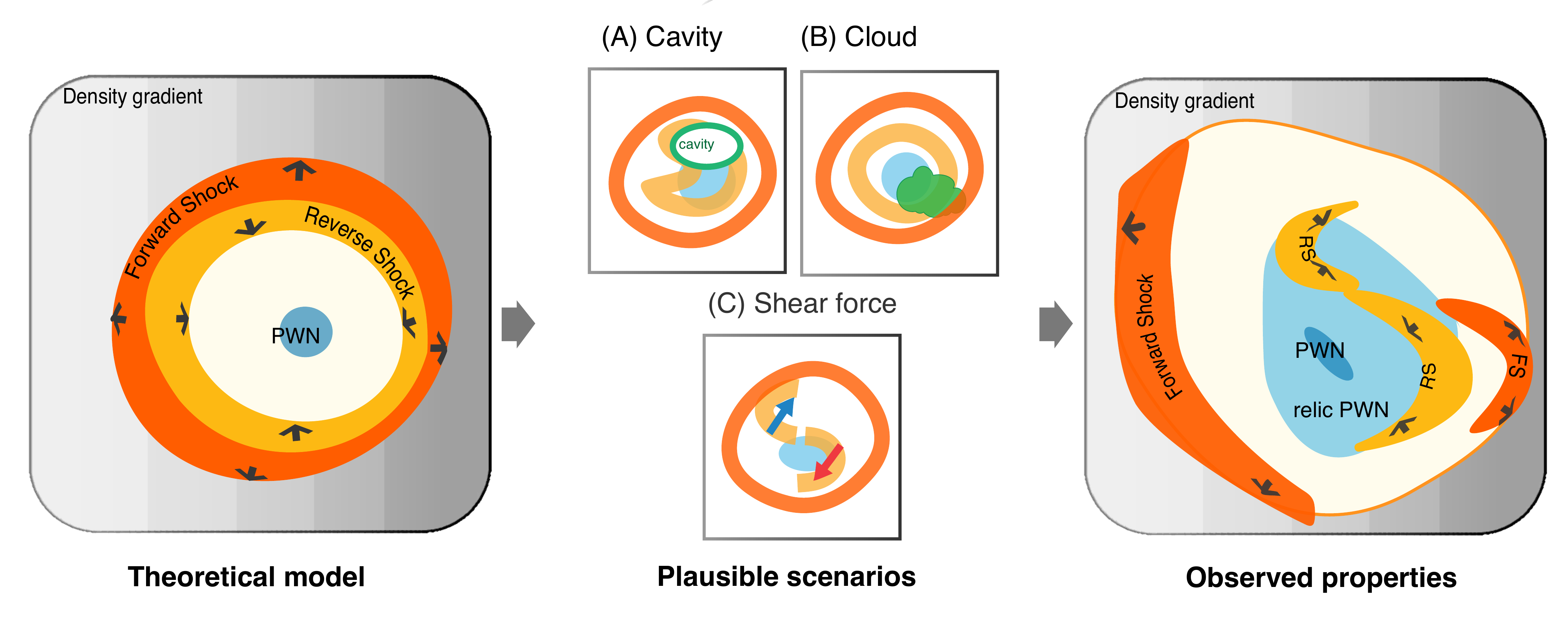}
  \caption{Plausible scenarios connecting the theoretical model and the observed properties of G309.8$-$2.6. The schematic illustration of physical model is adapted from \citet[][his Figure~3]{Slane2017}. The green line in the right panel shows the observed magnetic field orientation.} 
  \label{fig:sketch}
\end{figure*}

G309.8$-$2.6 shows complex morphology and properties, with forward shocks (FSs) and reverse shocks (RSs) appearing to be S-shaped, coexisting with the PWN and the relic PWN~(Figure~\ref{fig:sketch}, right panel). The general theoretical model of a PWN expanding into an SNR evolving in a medium with a density gradient, adapted from \citep[e.g.,][their Figure~3]{Slane2017}, is shown in Figure \ref{fig:sketch} (left panel). The density gradient provides a way to account for the different widths of the FSs. The eastern FS has a larger radius than the western FS, probably implying that the eastern region has a lower density than the western region, as illustrated by the gray background in Figure \ref{fig:sketch}. There could be several possible scenarios based on the observations that could connect the theoretical model and the observed properties. Because of the uncertainty of the RM foreground, we can not be sure whether there are RM reversals. Consequently, we can not derive the precise structure of the magnetic field. The ambient magnetic field and the magnetic field inside the SNR are also uncertain. Therefore, we did not include the magnetic field in Figure \ref{fig:sketch}. 

The system, which has an age of $\sim$$10^4$--$10^5$ yr, has likely evolved into the reverberation phase, and the reverse shock from the surrounding SNR collides with the PWN, compressing it and changing its internal magnetic and particle structure. As demonstrated by \citet{Ellison2005}, the reverse shock can produce considerable radio emission depending on the ambient or ejected magnetic field, implying strong amplification of the magnetic field. The reverse shock as well as its interaction with the PWN may strongly influence the observed morphology and synchrotron emission in the radio and X-ray bands.

The morphology can originate from the irregularities of the interstellar medium, such as cavities and clouds. If there is a pre-existing cavity (illustrated in Figure \ref{fig:sketch} inset (A)), which was not produced by the progenitor's stellar wind but by other massive stars, the interaction between the cavity wall and the RS could result in the asymmetry in Region II-1. We searched for a surrounding cavity in the H$\alpha$ and \textit{WISE} infrared data, but no clearly related structure was found. We also searched for nearby O/B-type stars \citep{Skiff2014,Chen2019}, but we did not find any definite associations. However, the absence of a cavity does not rule out complex prior stellar wind interactions. 

There might exist a cloud in Region II-2 (Figure~\ref{fig:sketch}, inset (B)), which is coincident with the H$\alpha$ emission in Figure~\ref{fig:infrared-ha}. 
A dense cloud can interact with the FS and still remain within the SNR interior, and then interact with the RS \citep{Jun1999}. The interaction could have also amplified the turbulent magnetic field, which contributed to the enhanced radio emission and caused depolarization, as observed in both cases. 

The overall morphology can also be shaped by the evolution of the SNR. If the pulsar has a large kick velocity, or if the progenitor of the SN is a runaway star \citep[e.g.][]{Meyer2022}, the ejected material in the same and opposite directions of the pulsar will have different speeds and thus experience different compressions. This will generate a shear force (Figure~\ref{fig:sketch}, inset (C)) that laterally displaces the two arcs and produces the S-shape in the relic PWN. The shear force will also drag the magnetic field and cause a reversal of the field direction, as indicated by the observed RM variations.

\section{Conclusion}\label{sec:conclusion}
We conduct a detailed study of G309.8$-$2.6 (the ``Salamander''). We present new radio continuum and polarization images from ASKAP EMU/POSSUM observations. We reprocess archival X-ray observations to obtain images, which trace PWN or gas heated by reverse shock. We retrieve archival H$\alpha$ images, which can trace radiative shock. We also retrieve infrared images, which indicate properties of the ambient interstellar medium. 

Combining multi-wavelength data, including infrared, H$\alpha$ and X-ray, we identify a complex system composed of an X-ray PWN powered by PSR~J1357$-$6429, a relic radio PWN, and a faint partial shell likely tracing the forward shock. These features, taken together, indicate that the source is of SNR origin.

The relic PWN shows a highly polarized S-shaped structure with an ordered magnetic field and a large-scale RM gradient or sign reversal across the PWN. We estimated the foreground RM to be 13$\pm$17 rad\,m$^{-2}$ based on the nearby pulsars. We also detect a radio tail extending from the pulsar, offset from the X-ray nebula, which likely traces the pulsar’s past proper motion. Assuming a distance of 1.8\,kpc, the relic PWN spans 12$-$18$d_{\rm 1.8}$ \,pc, and the full system reaches $R_{\rm s}=$ 20–42\,$d_{\rm 1.8}$ pc in diameter, consistent with an evolved system of age $10^4$--$10^5$\,yr.

The morphology, polarization signatures, and RM structure indicate that G309.8$-$2.6 is a middle-aged composite remnant in the reverberation phase, where the relic PWN is being reshaped by interactions with both the forward and reverse shocks. We also discuss several possible scenarios for the local environment. This system demonstrates the diagnostic power of radio polarimetry in tracing the evolution of PWNe and their host SNRs and highlights G309.8$-$2.6 as a valuable laboratory for studying the coupling between pulsar activity, shock dynamics, and the magneto-ionic interstellar medium.

\begin{acknowledgments}
We would like to thank the reviewer for the comments that have improved the presentation of the paper. We thank Dr. Wolfgang Reich for providing valuable comments. We thank Prof. Gavin Rowell for the valuable discussion.

This research has been supported by the National SKA Program of China (2022SKA0120101). 
W.-H. Jing is supported by the Scientific Research Fund Project of Yunnan Education Department (Project ID: 2025Y0015) and Scientific Research and Innovation Project of Postgraduate Students in the Academic Degree of Yunnan University (Project ID: KC-24248558). C.S.A. acknowledges funding from the Australian Research Council in the form of Australian Future Fellowship FT240100498.

This scientific work uses data obtained from Inyarrimanha Ilgari Bundara / the Murchison Radio-astronomy Observatory. 
We acknowledge the Wajarri Yamaji People as the Traditional Owners and native title holders of the Observatory site. 
The Australian SKA Pathfinder (ASKAP) is part of the Australia Telescope National Facility\footnote{\url{https://ror.org/05qajvd42}}, which is managed by CSIRO. 
Operation of ASKAP is funded by tfhe Australian Government with support from the National Collaborative Research Infrastructure Strategy. 
ASKAP uses the resources of the Pawsey Supercomputing Centre. 
Establishment of ASKAP, the Murchison Radio-astronomy Observatory, and the Pawsey Supercomputing Centre are initiatives of the Australian Government, with support from the Government of Western Australia and the Science and Industry Endowment Fund. 
The POSSUM project\footnote{\url{https://possum-survey.org}} has been made possible through funding from the Australian Research Council, the Natural Sciences and Engineering Research Council of Canada, the Canada Research Chairs Program, and the Canada Foundation for Innovation.
This work is based on data from eROSITA, the soft X-ray instrument aboard SRG, a joint Russian-German science mission supported by the Russian Space Agency (Roskosmos), in the interests of the Russian Academy of Sciences represented by its Space Research Institute (IKI), and the Deutsches Zentrum für Luft- und Raumfahrt (DLR). The SRG spacecraft was built by Lavochkin Association (NPOL) and its subcontractors, and is operated by NPOL with support from the Max Planck Institute for Extraterrestrial Physics (MPE). The development and construction of the eROSITA X-ray instrument was led by MPE, with contributions from the Dr. Karl Remeis Observatory Bamberg \& ECAP (FAU Erlangen-Nuernberg), the University of Hamburg Observatory, the Leibniz Institute for Astrophysics Potsdam (AIP), and the Institute for Astronomy and Astrophysics of the University of Tübingen, with the support of DLR and the Max Planck Society. The Argelander Institute for Astronomy of the University of Bonn and the Ludwig Maximilians Universität Munich also participated in the science preparation for eROSITA.
The eROSITA data shown here were processed using the eSASS software system developed by the German eROSITA consortium.
This research made use of Montage. It is funded by the National Science Foundation under Grant Number ACI-1440620, and was previously funded by the National Aeronautics and Space Administration's Earth Science Technology Office, Computation Technologies Project, under Cooperative Agreement Number NCC5-626 between NASA and the California Institute of Technology.
\end{acknowledgments}
\vspace{5mm}

\bibliography{0A}
\bibliographystyle{aasjournal}
\end{document}